\documentclass[sigconf]{acmart}
\usepackage{epstopdf}
\usepackage{subfigure}
\usepackage{booktabs} 
\usepackage{graphicx}
\usepackage{amsthm}
\usepackage{mathrsfs}
\usepackage{float}
\usepackage{multirow} 
\usepackage{amsmath} 
\usepackage{bm}

\usepackage{algorithm}
\usepackage{algcompatible}

\usepackage{array}
\usepackage{longtable}
\usepackage{multirow}
\usepackage{bigstrut}
\usepackage{makecell}
\usepackage{enumitem}
\usepackage{booktabs}  
\usepackage{threeparttable}  
\newtheorem{problem}{Problem}

\def\BibTeX{{\rm B\kern-.05em{\sc i\kern-.025em b}\kern-.08emT\kern-.1667em\lower.7ex\hbox{E}\kern-.125emX}}

\copyrightyear{2020} 
\acmYear{2020} 
\setcopyright{acmlicensed}\acmConference[CIKM '20]{Proceedings of the 29th ACM International Conference on Information and Knowledge Management}{October 19--23, 2020}{Virtual Event, Ireland}
\acmBooktitle{Proceedings of the 29th ACM International Conference on Information and Knowledge Management (CIKM '20), October 19--23, 2020, Virtual Event, Ireland}
\acmPrice{15.00}
\acmDOI{10.1145/3340531.3412015}
\acmISBN{978-1-4503-6859-9/20/10}

\begin{document}
\fancyhead{}

\title{Genetic Meta-Structure Search for Recommendation on Heterogeneous Information Network}

\author{Zhenyu Han $\ast$, Fengli Xu $\ast$, Jinghan Shi $\dagger$, Yu Shang $\ast$, Haorui Ma $\ast$, Pan Hui $\ddagger$, Yong Li $\ast$}

\affiliation{%
	\institution{$\ast$ Beijing National Research Center for Information Science and Technology (BNRist), \\Department of Electronic Engineering, Tsinghua University, Beijing, China, 100084}
	\city{}
	\state{}
	\country{}
	\postcode{}
}

\affiliation{%
	\institution{$\dagger$ Beijing University of Posts and Telecommunications, BUPT}
	\city{}
	\state{}
	\country{}
	\postcode{}
}

\affiliation{%
	\institution{$\ddagger$ University of Helsinki, The Hong Kong University of Science and Technology}
	\city{}
	\state{}
	\country{}
	\postcode{}
}

\email{hanzy19@mails.tsinghua.edu.cn, fenglixu2020@hotmail.com, superali@bupt.edu.cn, }
\email{{shang-y17, mhr17}@mails.tsinghua.edu.cn, pan.hui@helsinki.fi, liyong07@tsinghua.edu.cn}

\renewcommand{\shortauthors}{Zhenyu Han, et al.}

\begin{abstract}
In the past decade, the heterogeneous information network (HIN) has become an important methodology for modern recommender systems. To fully leverage its power, manually designed network templates, i.e., meta-structures, are introduced to filter out semantic-aware information. The hand-crafted meta-structure rely on intense expert knowledge, which is both laborious and data-dependent. On the other hand, the number of meta-structures grows exponentially with its size and the number of node types, which prohibits brute-force search. To address these challenges, we propose \emph{Genetic Meta-Structure Search} (GEMS) to automatically optimize meta-structure designs for recommendation on HINs. Specifically, GEMS adopts a parallel genetic algorithm to search meaningful meta-structures for recommendation, and designs dedicated rules and a meta-structure predictor to efficiently explore the search space. Finally, we propose an attention based multi-view graph convolutional network module to dynamically fuse information from different meta-structures. Extensive experiments on three real-world datasets suggest the effectiveness of GEMS, which consistently outperforms all baseline methods in HIN recommendation. Compared with simplified GEMS which utilizes hand-crafted meta-paths, GEMS achieves over $6\%$ performance gain on most evaluation metrics. More importantly, we conduct an in-depth analysis on the identified meta-structures, which sheds light on the HIN based recommender system design.
\end{abstract}

%
%



\maketitle

\section{Introduction}\label{sec:Introduction}
Heterogeneous information networks (HINs) \cite{zhao2017meta} have shown great potential for recommendation tasks due to the rich information provided by heterogeneous relations. In recommendation scenario, users can rate items, write reviews, and add new friends. Items also have different properties, which include brands, categories, origins et al.. The HIN is able to represent all relations above in a single graph, which provides much more information than traditional user-item interaction for recommendation tasks. 


To effectively extract semantic information, meta-path \cite{sun2011pathsim} is proposed. The meta-path defines a message-passing prototype that selects information according to specific semantics. Recently, the meta-structure \cite{huang2016meta,zhao2017meta} is introduced as a generalization of the meta-path, where the chain structure is replaced by a graph structure. In that case, meta-paths can be regarded as a special case of meta-structures. In the following paper, we refer meta-structures to represent both meta-paths and meta-structures as a more general form.

Although meta-structures are effective to mine the semantics in HINs, it is hard for us to design the meta-structures. First, we have little prior knowledge on meta-structure design. Moreover, the designed meta-structures are data-dependent and cannot be extended to other HINs. In that case, lots of human labor is needed to apply meta-structure based recommendation model on HINs.


In this work, we try to leverage the power of automated machine learning to automatically search meaningful meta-structures for recommendation. However, two challenges lie in the  automated HIN recommendation model. First, the number of meta-structures increases exponentially with nodes, which makes it impossible to brute-force search all of them on recommendation tasks. Second, the recommendation model needs to leverage multiple meta-structures for various semantics, which requires an effective method to combine information learned by searched meta-structures. Therefore, the varying combinations of meta-structures bring more difficulties on both the problem search space and model design.

To overcome these challenges, we propose Genetic Meta-structure Search, i.e., GEMS. Specifically, we design a genetic-algorithm-based model to automatically search meaningful meta-structures for recommendation tasks. The genetic algorithm generates new meta-structures for recommendation, while not all of the meta-structures are informative. To narrow down the search space, We set up several rules to make sure most of the generated meta-structures are meaningful for recommendation. Besides, we design a meta-structure predictor to pre-evaluate meta-structure performance before evaluating on real recommendation tasks. By filtering out badly performed meta-structures in advance, we can save computational resources for promising ones. To combine different semantics from meta-structures, we design an attention-based multi-view graph convolutional network (GCN) module that dynamically fuses learned embeddings for each meta-structure. Overall, GEMS achieves a performance gain of over $6\%$ compared with the simplified version that leverages hand-crafted meta-paths for recommendation tasks.

The contributions of this work can be summarized as follows:

\begin{itemize}[leftmargin=0.5cm]
\item To the best of our knowledge, this paper is the first work to explicitly search meta-structures for heterogeneous information networks. We transfer the genetic algorithm from traditional numerical optimization to meta-structure search problems. Our work saves human labor from complex meta-structure designs, resulting in a universal model for recommendation tasks in heterogeneous information networks.
\item We narrow down the problem search space by defining a reasonable meta-structure encoding method and mutation rules, alone with a meta-structure predictor to filter promising meta-structures before training on recommendation datasets.
\item We propose an attention-based multi-view GCN module to dynamically fuse information guided by different meta-structures for a better recommendation. 
\item We conduct extensive experiments to demonstrate the effectiveness of our proposed models while providing explainable results that may shed light on human-labored meta-structure design.
\end{itemize}

\begin{figure}[t]
 \centering
 \includegraphics[width=.45\textwidth,trim={0 0cm 0 0}]{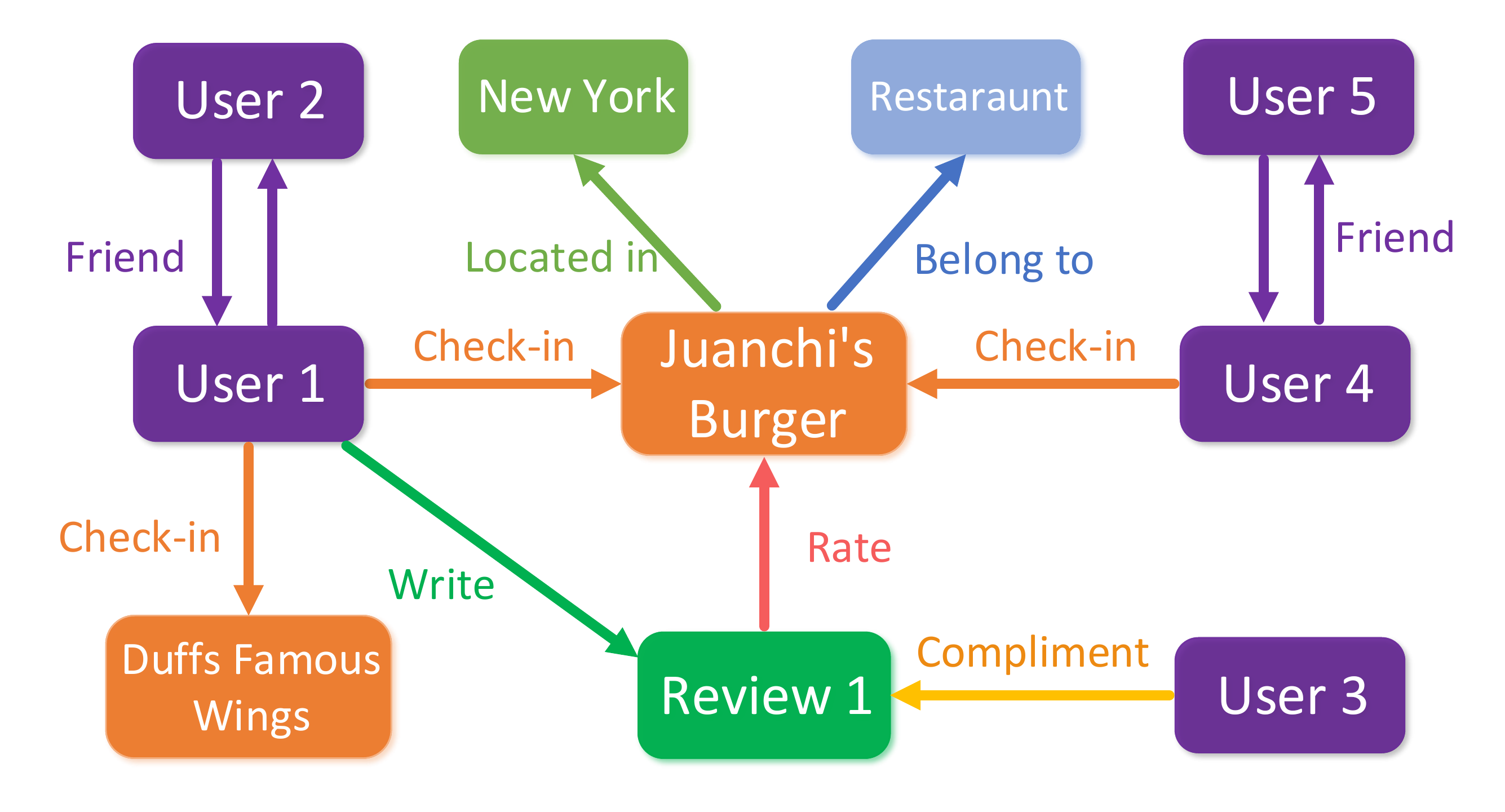}
 \caption{An example of HIN on Yelp platform.}
 \vspace{-0.2in}
 \label{fig:hin_yelp}
\end{figure}

\section{Preliminaries}

In real-world recommender systems, we often face complex relationships between different entities like Figure \ref{fig:hin_yelp} shows, where the relations of different entities are complex. To properly model these relations, heterogeneous information network is introduced.

\theoremstyle{definition}
\begin{definition}{\textbf{Heterogeneous Information Network}}~\cite{sun2011pathsim}.\\
Given a graph $\mathcal{G} = (\mathcal{V}, \mathcal{E})$, where $\mathcal{V}$ is the node set and $\mathcal{E}$ is the edge set. With a node mapping function $\phi(v):\mathcal{V}\rightarrow \mathcal{T}, \forall \ v\in \mathcal{V}$ and a relation mapping function $\psi(e): \mathcal{E}\rightarrow \mathcal{R}, \forall \ e\in \mathcal{E}$, where $\mathcal{T}$ is the node type set and $\mathcal{R}$ is the edge type set. If the number of node types  $|\mathcal{T}| > 1$ or number of relations types $|\mathcal{R}| > 1$, the directed graph $\mathcal{G} = (\mathcal{V}, \mathcal{E}, \mathcal{T}, \mathcal{R})$ is a heterogeneous information network.
\end{definition}

Heterogeneous nodes and edges make it more difficult to filter out useful information in HINs. As a result, researchers propose to mine HINs with hand-crafted semantic-aware network templates like meta-structures, which is defined as follows.



\theoremstyle{definition}
\begin{definition}\textbf{Meta-Structure} ~\cite{huang2016meta} Given a HIN $\mathcal{G} = (\mathcal{V}, \mathcal{E}, \mathcal{T}, \mathcal{R})$, a meta-structure $\mathcal{M} = (\mathcal{V}^{*}, \mathcal{E}^{*})$ is a sub-graph whose node $v^{*} \in\mathcal{T}$, edges $e^{*}\in\mathcal{R}$. The meta-structure $\mathcal{M}$ links a single source node $v^{*}_s$ and a single sink node $v^{*}_t$. 
\end{definition}

As mentioned in previous text, the design of meta-structure is complicated. Based on expert knowledge of recommender systems, we can design some collaborative filtering meta-structures like user-item-user-item, which depicts the phenomenon that people who have purchased the same items may have similar preferences. However, if we want to model more complex relationship which are usually data-dependent, we have to spend more energy into the design of meta-structures since previous experiences are very limited.

To save human labor from tedious meta-structure designing, we adopt automated machine learning paradigm to explicitly search meaningful meta-structures for recommendation. Specifically, we modify genetic algorithm to explore designed search space as well as choose promising meta-structures for optimization. Genetic algorithm \cite{goldberg2006genetic} is a search heuristic that is inspired by natural evolution theory. Genetic algorithm maintains a \emph{population} of models, each model containing multiple \emph{genes} that encode the search space instances. Each model is called an \emph{individual}. During each generation, there are possibilities for each individual's genes to \emph{mutate} and \emph{crossover} to explore the search space, then the environmental evaluator will score each individual and \emph{eliminate} individuals who perform badly. Finally, the remain individuals \emph{reproduce} according to the evaluated score to increase the percentage of promising genes. Finally, the recommendation problem in HINs can be defined as follows.

\theoremstyle{Problem}
\begin{problem}{\textbf{Recommendation in HINs.}}
Given a HIN $\mathcal{G}$ with user's purchase records dataset $\mathcal{D}=\left\{<v_u, v_i>\right\}$, where $v_u, v_i$ stands for user id and item id accordingly. For each user, we leverage meta-structure $\mathcal{M}$ to filter new interactions as side-information to generate a ranked list of items that are of interest to the user.
\end{problem}

\section{Method}\label{sec:Method}
Effective meta-structure design takes lots of human effort, which inspires us to leverage automated machine learning paradigm to search promising meta-structures for recommendation. However, the search space of meta-structures grows exponentially with its size, while the varying combination of node type $\mathcal{T}$ in HINs further enlarges the search space of meta-structures. Another problem emerges during the utilization of meta-structures. Each meta-structure corresponds to a semantic-aware feature extracted from HINs. To fully harness the power of HINs, recommendation models usually adopt multiple meta-structures. How to determine their importance and combine them together in a single model is another important issue in HIN recommendation.

In GEMS model, we propose a novel genetic framework to automatically design effective meta-structures. Besides, we carefully design a set of mutation rules and evaluation modules to further narrow down the search space. As for the second challenge on leveraging information of multiple meta-structures, we introduce an attention based multi-view GCN module for recommendation. 

\subsection{Genetic Framework}\label{subsec:genetic_ops}

\begin{figure}[t]
 \centering
 \includegraphics[width=.5\textwidth,trim={0 1.5cm 0 0}]{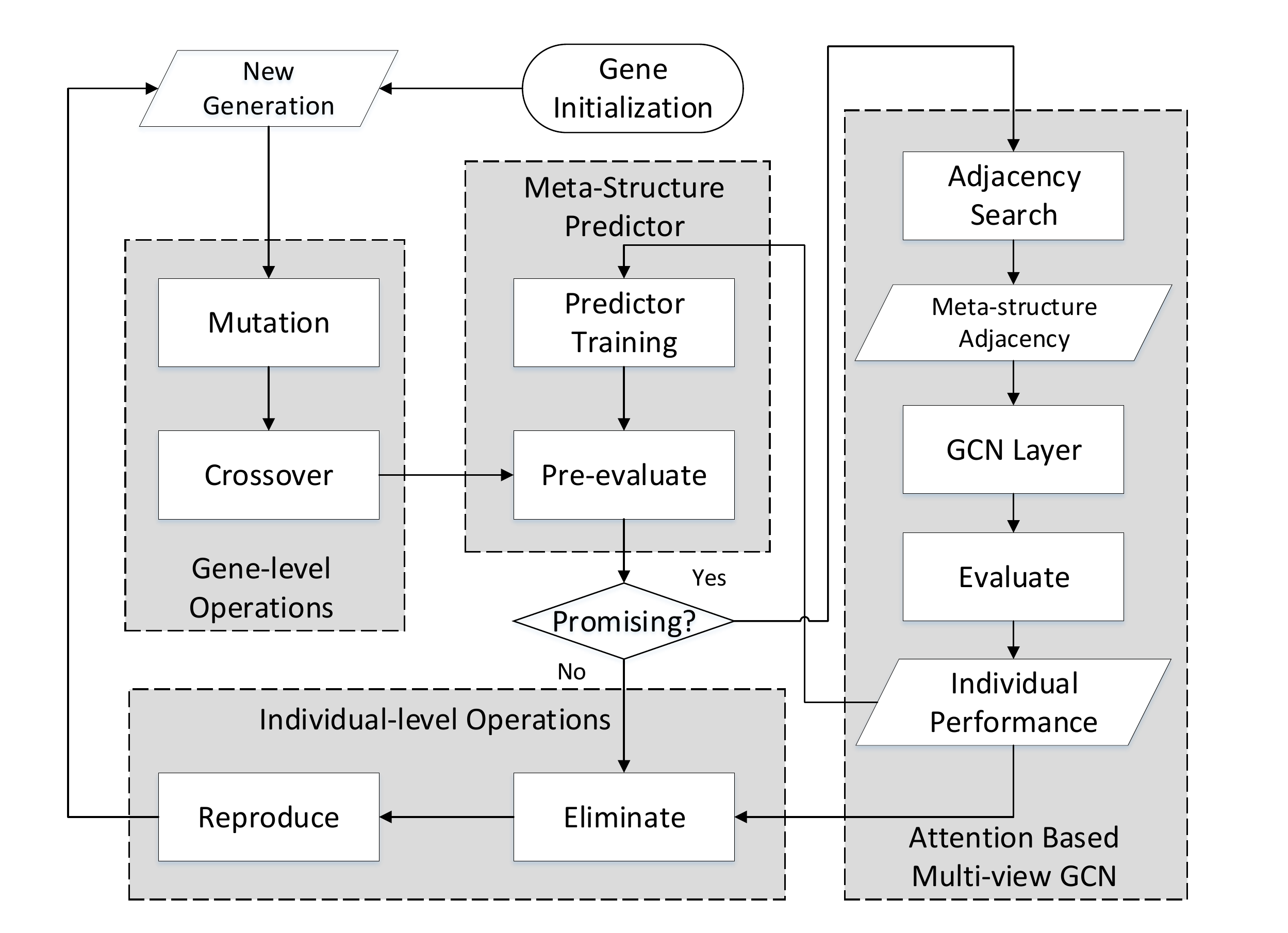}
 \caption{Flow chart of GEMS.}
 \vspace{-0.1in}
 \label{fig:gems}
\end{figure}

The genetic algorithm maintains a \emph{population} as a collection of \emph{individuals}. Each individual possesses several genes as its unique features. In our HIN recommendation scenario, an individual is a recommendation model with selected meta-structures. To leverage the power of genetic algorithms, we need to encode these meta-structures into genes as the problem search space. Such an encoding method should satisfy the property that every meta-structure can be represented by the encoded gene. In that case, we encode the meta-structure by an ordered node type list $T_M$ and an adjacency matrix $\bm{A}_M$ according to linkages in the meta-structure.

We first connect the meta-structure search with genetic algorithm by encoding meta-structure into gene representation. As shown in Figure \ref{fig:gems}, all the genes of individuals are \emph{mutated} and then \emph{crossover} with each other to generate new genes or new combinations of genes. These two operations are introduced to explore the problem search space. Then each individual will be pre-evaluated by the meta-structure predictor of GEMS, which aims to judge the performance of individuals by their gene combinations. For promising individuals with high pre-evaluated scores, they will be evaluated on recommendation tasks by the attention based multi-view GCN module. For individuals with low pre-evaluated scores, we directly use this predicted score as their performance. To select promising meta-structures, in genetic algorithm we \emph{eliminate} individuals by their performance. That is, we preserve individuals with high performance while delete badly performed ones. In that case, promising individuals and meta-structures therein will be preserved. Further, the remaining individuals will \emph{reproduce} according to their performance to construct a new generation. The reproduce process generate duplicates of promising individuals to maintain the population size.

To understand how these genetic operations work in the HIN recommendation scenario, we divide them into two categories. At the gene level, \emph{mutation} will change the structure of genes, which results in new meta-structures. According to the representation of genes, mutation can be divided into two types: add/delete edges as well as add/delete nodes. Edge mutation only affects the adjacency matrix, which randomly flips elements. Node mutation will add/delete nodes in the gene, which needs to add/delete elements in both the type list and the adjacency matrix. For example, if we need to delete a specific node, it will be removed from type list $T_M$, then the corresponding row and columns will also be removed from $\bm{A}_M$. On the other hand, if we need to add a new node, first we randomly choose a node type and append it at the end of $T_M$. Then, a new row and new column will also be appended to $\bm{A}_M$. Elements of the new row and new column are randomly set. Meanwhile, \emph{crossover} operation exchanges genes among individuals to strengthen gene circulation. Gene itself will not be modified in crossover operation. In conclusion, mutation and crossover operations enable GEMS to explore all possible search space of meta-structures effectively.

At the individual level, \emph{elimination} and \emph{reproduction} operations will guide the optimization process to find promising meta-structures. Genetic algorithm obsoletes individuals who perform badly on recommendation tasks, and the corresponding genes will be deleted. The remaining individuals generate their gene copies to form new individuals according to their performance, which makes promising genes prosper in the whole population. In that case, GEMS evolves to leverage better meta-structures for recommendation. 

\vspace{-0.05in}
\subsection{Search Space Optimization}\label{subsec:search_space}
As mentioned before, the search space of meta-structures is huge. Particularly, there are three factors that enlarge the search space. First, the node type list $T_M$ has different combinations due to the heterogeneous nodes in HIN. Given a HIN $G = (\mathcal{V}, \mathcal{E}, \mathcal{T}, \mathcal{R})$, the number of node types is defined as $m:=\vert \mathcal{T}\vert$. For a meta-structure containing $n$ nodes, there exists $m^n$ type lists. Second, the connection between meta-structure nodes is another important factor that enlarges the search space. Even if the node type list $T_M$ is fixed, there are still $2^{n\times n}$ adjacency matrices that results in different genes. Third, since the recommendation model leverages multiple meta-structures at the same time, different combinations of meta-structures will further extend the required search space. If the recommendation model adopts $k$ meta-structures, in the worst case the problem search space is $\binom{m^n \times 2^{n\times n}}{k}$.

The huge search space makes it impossible to find optimal meta-structures for recommendation. According to the above analysis, we propose three constraints to avoid meaningless meta-structures during the search process. First, based on the domain knowledge of recommender system, we impose several constraints on meta-structure genes to eliminate infeasible choices. Second, we set up three mutation rules to make sure that new genes produced by the mutation process introduce meaningful information for recommendation. Third, we design a meta-structure predictor to reduce the number of individuals that will be evaluated on recommendation tasks.

\subsubsection{Meta-Structure Encoding}

Based on the expert knowledge of recommendation task, we can narrow down the encoding space of meta-structure with the following constraints. We make the following observations of HIN based recommender systems. 

\begin{itemize}[leftmargin=0.5cm]
    \item Relations in HIN are constrained. Take Yelp as an example, \emph{category (A)} nodes can only link with \emph{business (B)} nodes, while linkages with \emph{user (U)} nodes are not allowed.
    \item Models leveraging HIN for recommendation usually concern the source and the sink nodes of meta-structures but not the whole path. For example, in \cite{wang2019heterogeneous,zhao2017meta} meta-structures perform as semantic filters, where new interaction matrices under corresponding semantics are constructed to provide side information for recommendation.
\end{itemize}

Based on the first observation, during the encoding process, we consider possible relations in the HIN by presetting forbidden links in adjacency matrix $\bm{A}_M$ as $-1$. In that case, we exclude meta-structures that are not practical for specific HIN. According to the second observation, we focus on the source node and sink node only. We refer these two nodes as target nodes in the following text. We fix the first two nodes of $T_M$ as the target nodes. For example, in Yelp dataset, the target nodes are \emph{user (U)} and \emph{business (B)} accordingly, since we need to recommend businesses for users. In recommendation scenario, the influence between users and items is mutual, which means the link between the target nodes is bi-directional. In that case, we can omit the direction of meta-structure, resulting in an upper triangular adjacency matrix. 

\begin{figure}[t]
 \centering
 \includegraphics[width=.4\textwidth,trim={0 1.2cm 0 0}]{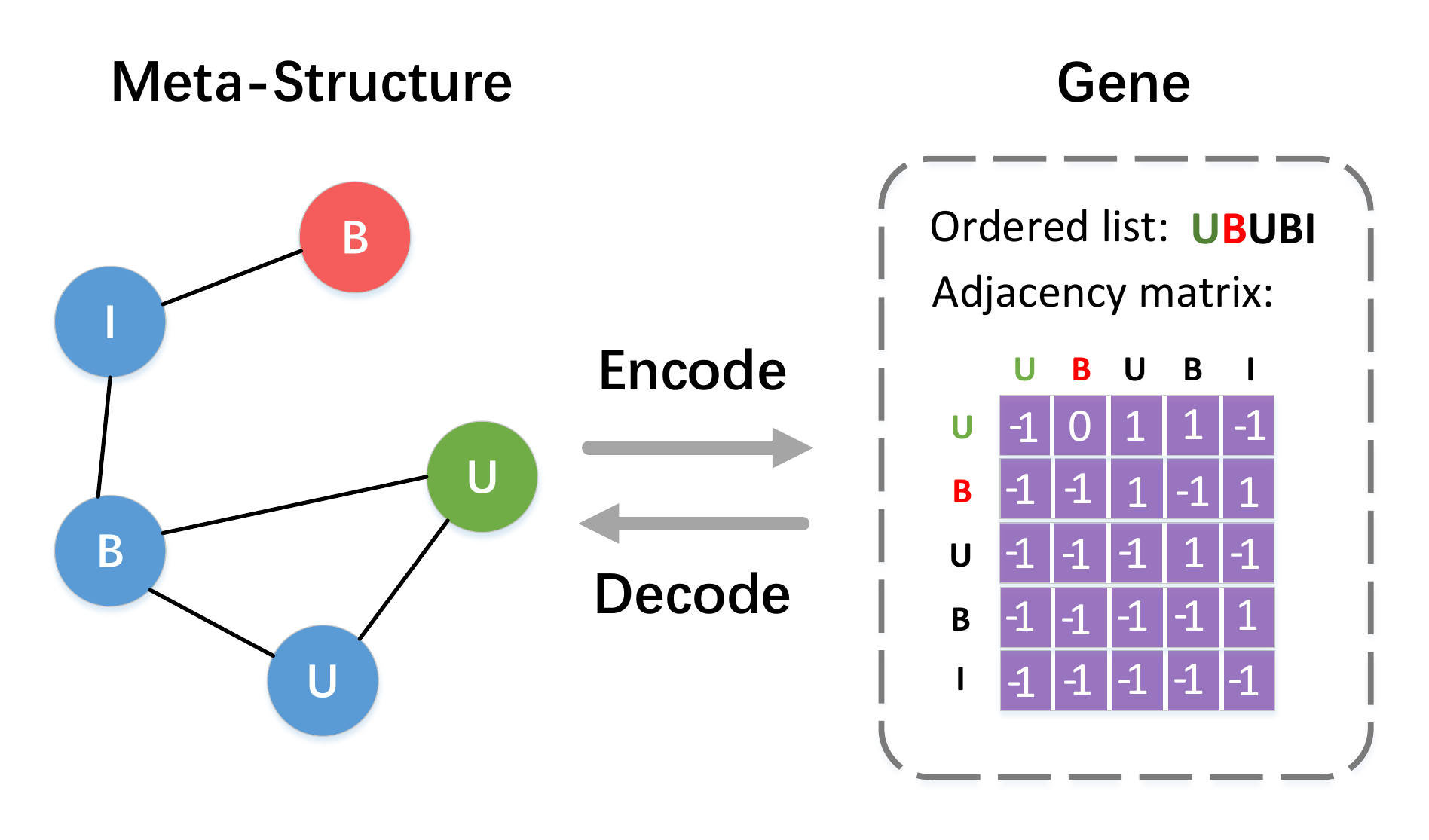}
 \caption{An illustrating example of meta-structure encoding in GEMS.}
 \vspace{-0.1in}
 \label{fig:encoding}
\end{figure}

Figure \ref{fig:encoding} demonstrates a meta-structure with its encoded gene in Yelp dataset. The source node is set as \emph{user (U)}, which is colored as green. The sink node is \emph{business (B)}, which is colored as red. According to the optimized encoding method, most elements in our designed adjacency matrix is $-1$, which greatly reduces the problem search space. In this illustrative example, the number of remaining non -1 positions is reduced from 25 to 7, which will be helpful to narrow down the problem search space.

\subsubsection{Mutation Rules}

During evolution, new genes will be generated by the mutation operation. However, many mutated genes are meaningless for recommendation. We introduce three mutation rules to ensure most of the new genes will exert influence on recommendation. 

\begin{figure*}[t]
 \centering
    \subfigure[Non-exist links]{
        \label{fig:muta_r1}
        \includegraphics[height=0.15\textwidth]{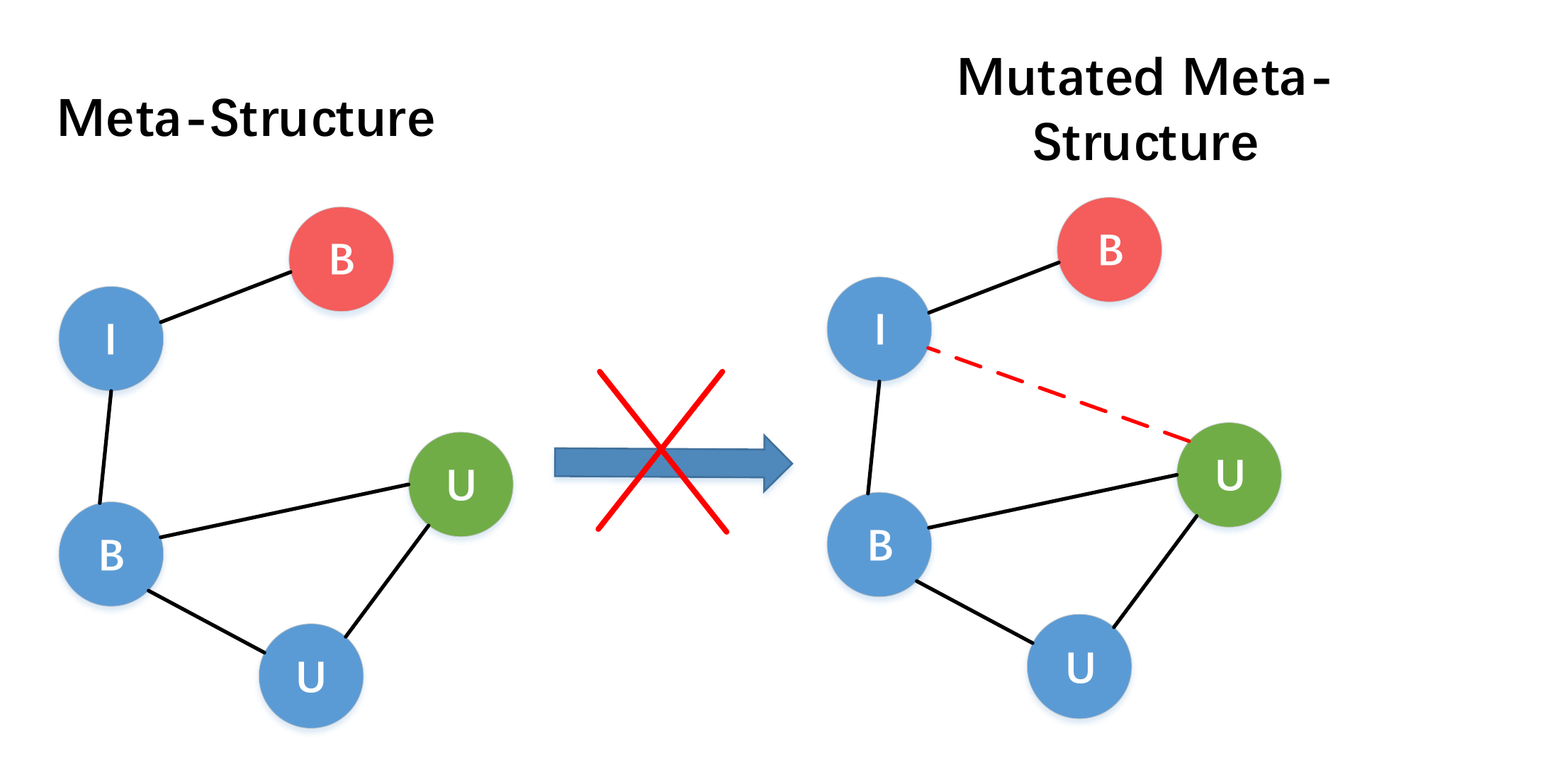}}
    \vspace{-0.1in}
    \subfigure[Constant information loops]{
        \label{fig:muta_r2}
        \includegraphics[height=0.15\textwidth]{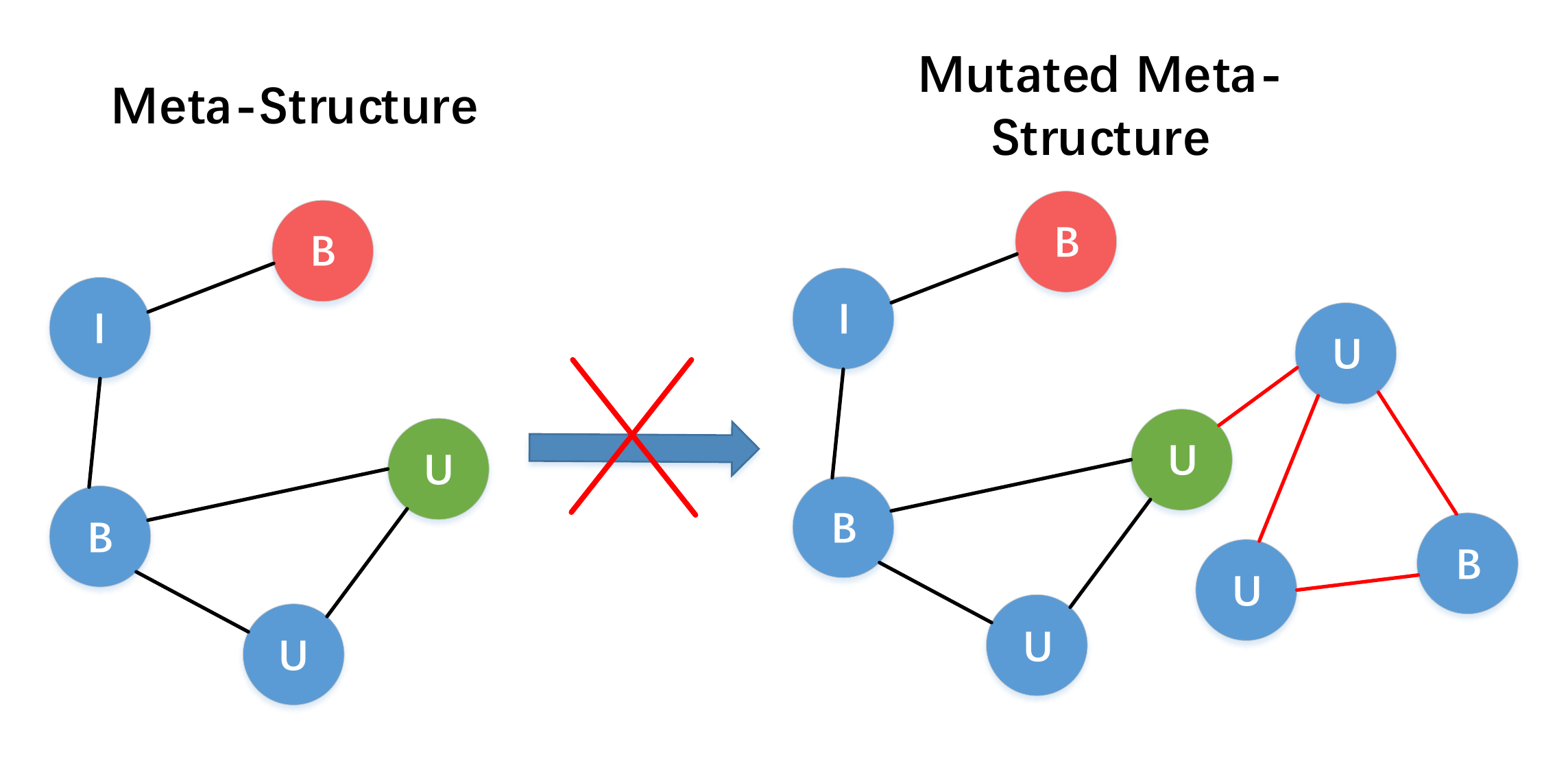}}
    \vspace{-0.1in}
    \subfigure[Side-branches]{
        \label{fig:muta_r3}
        \includegraphics[height=0.15\textwidth]{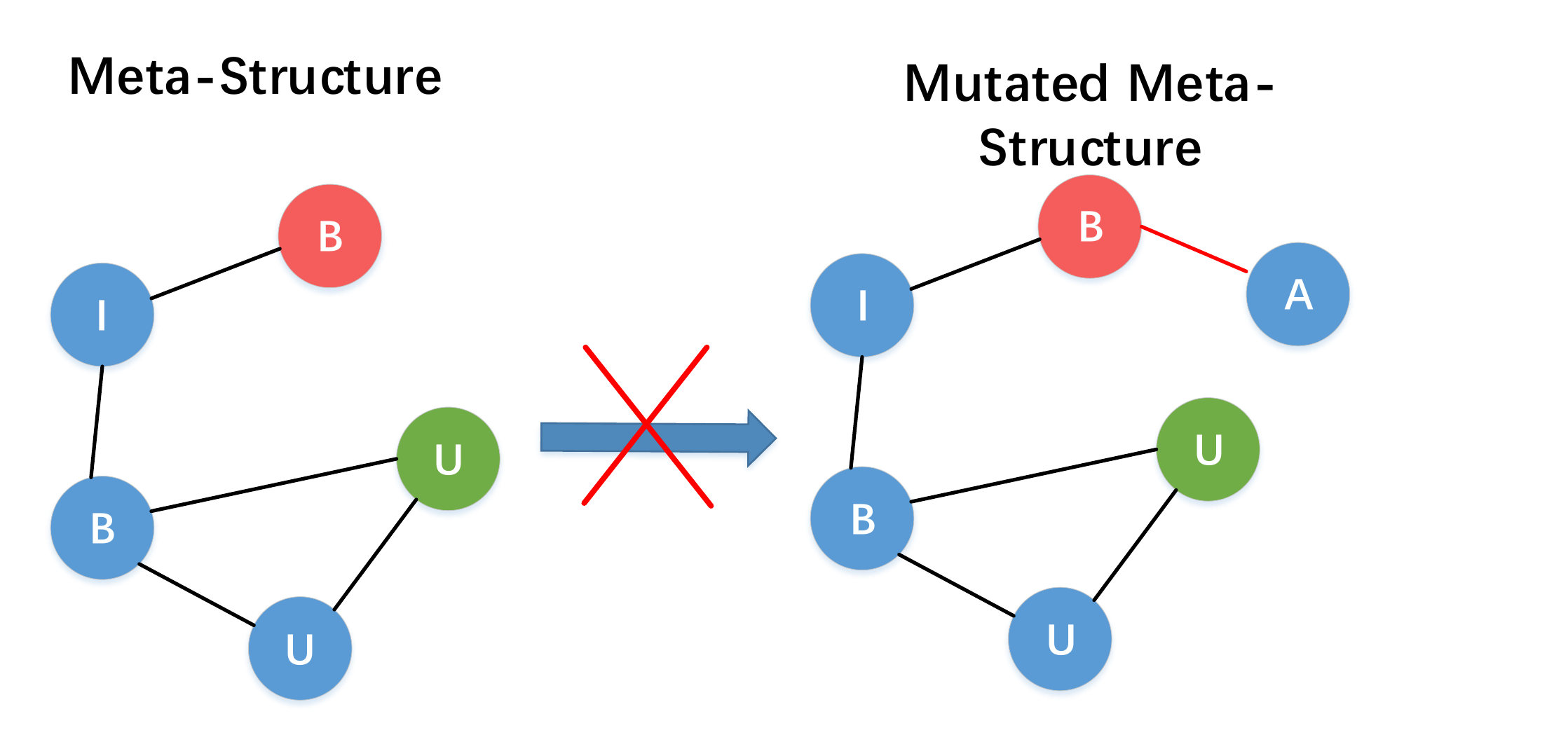}}
 \caption{Examples of mutation rule violations.}
 \label{fig:mutation_rules}
\end{figure*}

\begin{itemize}[leftmargin=0.5cm]
\item Avoid non-exist links: According to the definition of the meta-structure gene, during each mutation we will not flip elements of $-1$. This rule promises the effectiveness of new genes so that all mutated meta-structure are available in HINs. As shown in Figure \ref{fig:muta_r1}, the mutated graph connects \emph{user (U)} and \emph{city (I)}, which is not allowed in this scenario.
\item Avoid constant information loops: In recommendation scenario, new interactions are extracted according to meta-structures. If the mutation does not affect the paths between target nodes, there will be little difference in searched adjacency matrix. Since the following GCN module is based on this adjacency, the recommendation performance will not receive much performance gain. Based on this understanding, we have the second rule that the paths between target nodes should be changed during mutation. In Figure \ref{fig:muta_r2}, the mutation does not introduce new paths from the green \emph{user (U)} to the red \emph{business (B)}, which is considered invalid.
\item Cut off side-branches: After mutation there will be side-branches in the meta-structure, e.g. a single node that only has one edge with all other nodes. These side-branches will not affect the effectiveness of meta-structures since they do not provide any information. In that case, after the mutation we delete all the side-branches to further narrow down gene numbers that need to be evaluated in the following module. Figure \ref{fig:muta_r3} demonstrate such side-branches. Since each \emph{business (B)} has a corresponding \emph{category (A)} connection, these side-branches are useless for recommendation.
\end{itemize}

\subsubsection{Meta-Structure Predictor}

Different combinations of meta-structures will greatly affect recommendation performance. The meta-structure predictor tries to infer recommendation performance by meta-structure combinations for each individual. Introducing the meta-structure predictor to filter promising individuals can save lots of computational resources and explore more meta-structures in limited time. In the worst case, it can remember the performance of evaluated meta-structures, which is also helpful for speed-up search process. Here, we build a small GCN network according to the meta-structures (rather than the adjacency matrix it corresponds). Each kind of node type has an embedding vector, and the GCN layer is built on connections defined by the meta-structure. The meta-structure predictor is trained by the real recommendation 
performance of each individual. We normalize the metrics into $-1 \sim 1$ to generate higher accuracy predictions. The meta-structure predictor adopts MSE loss function for training.

With the above methods for search space optimization, the GEMS model can explore more meta-structures and select meaningful ones for recommendation purpose in reasonable time.

\subsection{Attention Based Multi-view GCN Design}\label{subsec:gcn}
To leverage different semantic information in HIN, recommendation models need to adopt multiple meta-structures to extract corresponding knowledge. In GEMS, we design a multi-view GCN architecture to fuse semantic information guided by meta-structures.

\begin{figure}[t]
 \centering
 \includegraphics[width=.45\textwidth,trim={0 0cm 0 0}]{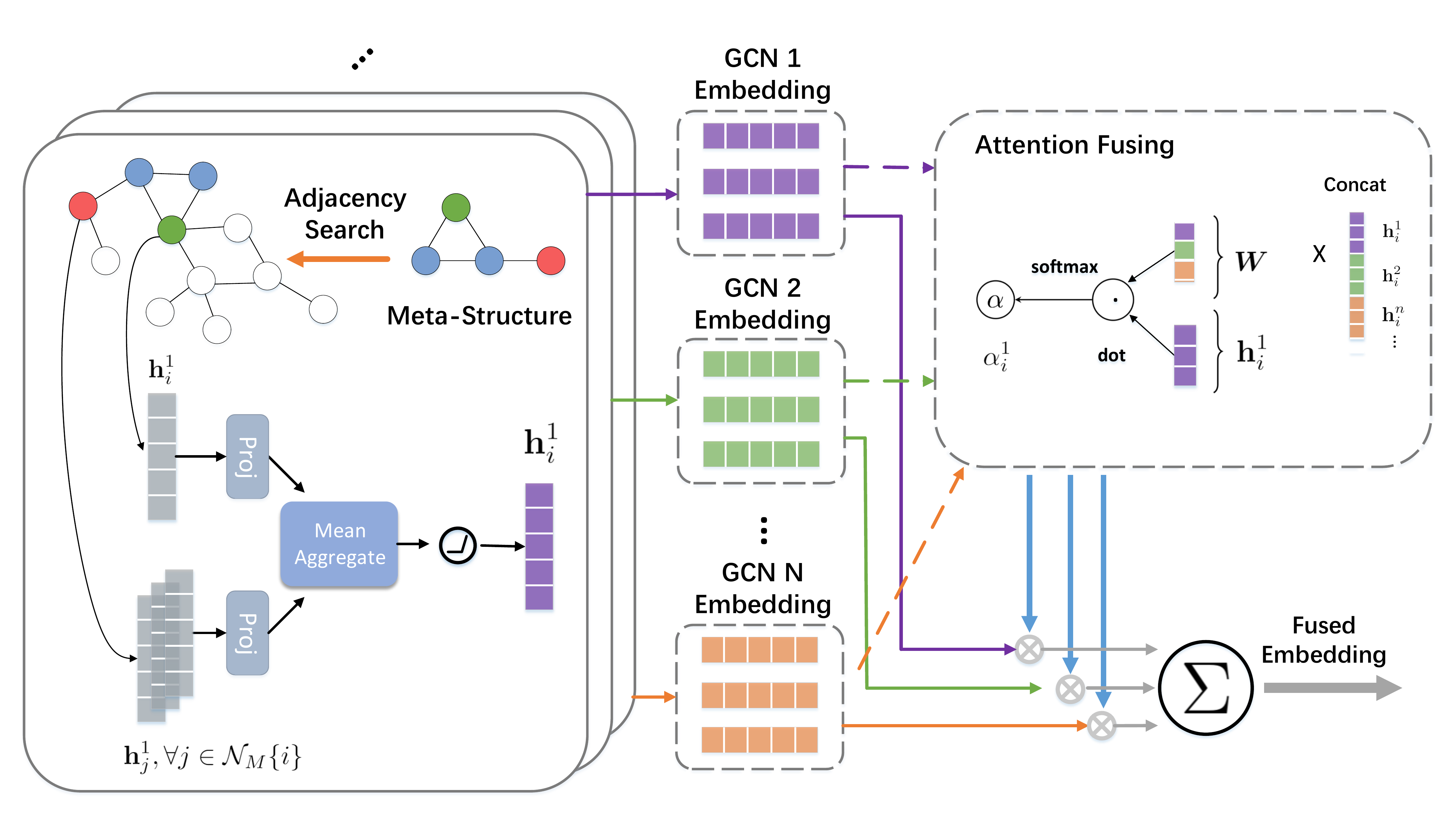}
 \caption{Attention based Multi-view GCN.}
 \label{fig:attention_fuse}
\end{figure}

Figure \ref{fig:attention_fuse} demonstrates the architecture of proposed multi-view GCN on \emph{user (U)} nodes. For each individual, GEMS builds multiple independent GCN layers by different meta-structure adjacency matrices. These GCN layers will learn embedding vectors that have different semantic meanings on HIN recommendation tasks. To fully leverage the combinational effect on semantics, we fuse these embedding vectors in following attention fusing module. Finally the fused embeddings are used to calculate similarities between user-item pairs for recommendation tasks.

During the construction of meta-structure adjacency, sampling is needed to control the number of linkages in the adjacency matrix since the neighbor size goes quickly with the depth of the meta-structure. To ensure fairness during sampling, the search process begins from the node type which has the biggest degree in the HIN. We denote the adjacency matrix corresponding with meta-structure $\mathcal{M}$ as $\mathcal{N}_M$, and the neighbor set of node $i$ is $\mathcal{N}_M\{i\}$.

According to different adjacency matrices defined by meta-structures, we construct multiple interaction graphs. Upon these interaction graphs, GEMS trains different GCN layers to capture corresponding semantic information on HIN. It should be noted that, the meta-structure can connect two far-away nodes, which enlarge the receptive field of GCN. In that case, we adopt a single layer GCN for each meta-structure to reduce the computational overhead.


To fuse the embeddings learned by different GCNs together, we propose an attention based embedding fusion mechanism to dynamically assign the importance weight for every meta-structures of the individual.

Each GCN layer generates an embedding as $\textbf{h}^\star \in \mathbb{R}^{d}$ for target nodes, where $\star$ represents different GCN layers and $d$ is the embedding size. We concatenate $n$ embeddings together, then apply a transformation matrix $\bm{W} \in \mathbb{R}^{d\times nd}$ to map the concatenated embedding into the same size of GCN embedding. This embedding serves as query vectors $\textbf{q}_i\in\mathbb{R}^{d}$, that is

\begin{equation}
\textbf{q}_i = f\big(\bm{W}(\textbf{h}^1_i \oplus\textbf{h}^2_i\dotsc\oplus\textbf{h}^n_i)\big),
\end{equation}
where $f(\cdot)$ denotes a nonlinear activation function, and $\oplus$ denotes the concatenation operation. 

With the query vector $\textbf{q}_i$, we can assign attention weights for different meta-structure embeddings. For meta-structure embedding $\star \in \{1,2,\dots,n\}$, we have attention weight $\bm{\alpha_i}$ expressed as follows:

\begin{equation}
    \bm{\alpha}_i = \mathrm{softmax}(\textbf{h}^{\star}_i \cdot \textbf{q}_i).
\end{equation}

The fused embedding is computed as the attention coefficient weighted sum over the embeddings derived from different meta-structures, which is obtained as follows:
\begin{equation}
    y_i = \sum_{\star=1}^{n} \alpha^\star_i \textbf{h}^{\star}_i.
\end{equation}
Finally, the inner product of fused embeddings is used to evaluate the similarity between users and items.

According to the embedding fusing module, the GEMS model combines different semantics from meta-structures for individuals.

\subsection{Training and Learning}
In GEMS, the attention based multi-view GCN is trained in a supervised learning approach with the purchase record of dataset $\mathcal{D} = \{<v_u,v_i>\}$, where $v_u$ stands for users ans $v_i$ stands for items. The specific meaning of the item depends on the dataset. The recommendation likelihood is calculated as follows:
\begin{equation}
    z(v_u,v_i) = \sigma \big(y_{v_u} \cdot y_{v_i} \big),
\end{equation}
where $y_{v_u}, y_{v_i}$ are fused embeddings of the GCN backend, and $\sigma(\cdot)$ is the sigmoid activation function. 

For any positive purchase record in dataset $\mathcal{D}$, we sample a pre-defined number of negative items $v_n$ according to the occurrence frequency of items. We adopt a max-margin based ranking loss function to train the GCN backend as follows:
\begin{equation}
J_{\mathcal{G}}(v_u, v_i) = \mathbb{E}_{v_n}\mathrm{max}\left\{0,\ z(v_u, v_n)-z(v_u, v_i)+\Delta \right\},
\end{equation}
where $\Delta$ denotes the hyper-parameter of pre-defined margin. The intuition is to train the model to predict the positive samples with a higher likelihood by a pre-defined margin than negative items.

Individuals are fully independent after crossover operation, and meta-structure adjacency search and GCN training can be fully paralleled. This enables us to leveraging the high concurrency ability of modern computing hardware to speed-up model convergence. To better understand the procedure of GEMS, we provide a detailed description as shown in Algorithm \ref{alg:gems}.

\begin{algorithm}
\caption{: Genetic Meta-Structure Search}\label{alg:gems}
\begin{algorithmic}[1]
\REQUIRE{HIN $G=(\mathcal{V}, \mathcal{E})$, Dataset $\mathcal{D}=\left\{<v_u, v_i>\right\}$, node features $\left\{\mathbf{x}_v, \forall v\in V \right\}$, epoch number $e$}
\ENSURE 
\STATE Initialize  $genePool$ for the $population$
\STATE Initialize meta-structure predictor $predic(\cdot)$
\FOR{$epoch=1$ to $e$} 
\STATE /*Mutate and Crossover*/
\STATE $genePool$ = Mutate$(genePool)$ 
\STATE $genePool$ = Crossover$(genePool)$ 
\FOR{$individual$}
\STATE Assign $individualGenes$ from $genePool$
\STATE /*Using predictor to filter promising individuals*/
\IF{$predic(individualGenes)<threshold$} 
\STATE $metric$ = $predic(individualGenes)$
\STATE continue
\ELSE
\FOR{$gene$ in $individualGenes$}  
\STATE /*Adjacency Search*/
\STATE $\mathcal{N}_{gene} = \mathrm{AdjSearch}(gene) $
\ENDFOR 
\STATE /*Train the GCN and preform recommendation task*/
\STATE $metric$ = GCN($\mathbf{x}_v,\mathcal{D},\mathcal{N}$)
\ENDIF
\ENDFOR 
\STATE /*Eliminate and Reproduce*/
\STATE $population$ = Eliminate($population,metrics$)
\STATE $population$ = Reproduce($population,metrics$)
\STATE Train meta-structure predictor by $(individual, metric)$ pairs

\ENDFOR 
\end{algorithmic}
\end{algorithm}

\section{Experiments}\label{sec:Experiments}
\subsection{Datasets and Baselines}
We evaluate our proposed solution GEMS as well as competitive baselines on three widely used real-world recommendation datasets with rich heterogeneous information, Yelp\footnote{\url{https://www.yelp.com/dataset challenge}}, Douban Movie\footnote{\url{https://movie.douban.com/}} and Amazon\footnote{\url{http://jmcauley.ucsd.edu/data/amazon/}}. Yelp is a platform for users to rate local businesses and share photos for others' reviews. Douban Movie is a well-known social media network in China for users to rate, share, and comment movies. Amazon is one of the biggest e-commerce platforms with global operations. These three datasets have different densities of records and semantic information, which guarantees the reliability of experiments.

The basic statistics and relations of these datasets are available in Table \ref{tab:dataset}. Douban Movie is a dense dataset, whose rating matrix achieves a density of $0.630\%$. In the meantime, it preserves rich information for HIN recommendation up to contains 6 relations. As for the Amazon dataset, user-user relationship does not exist. Thus, user nodes can only link with item nodes. The Amazon dataset represents traditional recommendation information, where the relation types are relatively limited. Yelp dataset has a sparse rating matrix with a density of $0.086\%$. However, it contains three relations for user nodes and two relations for business nodes, which preserves rich semantics for HIN recommendation.


\begin{table}[t]
 \small
	\centering
	\caption{The basic information of evaluation datasets.}
	\begin{tabular}{|m{1cm}<{\centering}||m{3.2cm}<{\centering}|m{0.8cm}<{\centering}|m{0.8cm}<{\centering}|m{0.9cm}<{\centering}|}
	    \hline
	    Dataset& Relations (A-B)& $\#$A & $\#$B & $\#$ A - B\\ 
	    \hline
	    \hline
	    \multirow{5}{*}{Yelp}&	{\bf User-Business (U-B)}&{\bf 16239}&{\bf 14284}&{\bf 198397}\cr	\cline{2-5}  &
	    User-User (U-U)&16239&16239&158590\cr	\cline{2-5}&
	    User-Compliment (U-O)&16239&11&76875\cr	\cline{2-5}&
	    Business-City (B-I)&14284&47&14267\cr	\cline{2-5}&
	    Business-Category (B-A)&14284&511&40009\\
	    \hline
		\hline
        
        \multirow{6}{*}{\makecell[c]{Douban\\Movie}}&{\bf User-Movie (U-M)}&{\bf 13367}&{\bf 12677}&{\bf 1068278}\cr	\cline{2-5}&
        User-Group (U-G)&13367&2753&570047\cr	\cline{2-5}&
        User-User (U-U)&13367&13367&4085\cr	\cline{2-5}&
        Movie-Actor (M-A)&12677&6311&33587\cr	\cline{2-5}&
        Movie-Director (M-D)&12677&2449&11276\cr	\cline{2-5}&
        Movie-Type (M-T)&12677&38&27668\\ 
		\hline
		\hline
		
		\multirow{4}{*}{Amazon}&{\bf User-Item (U-I)}&{\bf 6170}&{\bf 2753}&{\bf 195791}\cr	\cline{2-5}&
		Item-View (I-V)&2753&3857&5694\cr	\cline{2-5}&
		Item-Category (I-C)&2753&22&5508\cr	\cline{2-5}&
		Item-Brand (I-B)&2753&334&2753\\ 
		\hline
	\end{tabular}
	\label{tab:dataset}
\end{table}

In order to show the performance gain of our proposed GEMS model, we compare it with eight state-of-the-art baselines in recommendation tasks. Specifically, they include traditional matrix factorization based models (NMF, BMF, Metapath MF, and SVD++), GCN-based models (PinSAGE and GAT), alone with recent models that leverage meta-structures in HINs (HAN and FMG). Recently, new HIN approaches inspired by transformer network like GTN \cite{yun2019graph} and HGT \cite{hu2020heterogeneous} emerged. In these model there are no explicitly defined meta-structures, and meta-structures are only used to explain the learnt importance of different adjacency matrices. From this perspective, they are different from other models and not included in the baseline comparison.

\begin{itemize}[leftmargin=0.5cm]
\item{NMF \cite{lee1999learning}:} A matrix factorization model that results in non-negative matrices, which represents additive features of users and items.
\item{BMF \cite{koren2009matrix}:} A matrix factorization model that contains bias term for each user and item to better depict personal characteristics.
\item{Metapath MF \cite{vahedian2016meta}:} In Metapath MF, new interaction matrices are generated based on meta-structures defined in Table~\ref{tab:metapath}, then train multiple basic MF model accordingly. The learned embeddings are averaged as fused embedding, then the embedding is processed by a linear transformation. Finally, we perform inner product on the output embeddings as MF to generate recommendation scores.
\item{SVD++ \cite{koren2008factorization}:} Enhanced singular value decomposition algorithm for recommendation.
\item{PinSAGE \cite{ying2018graph}:} A GCN model on the homogeneous graph that samples fixed size of neighbors during aggregation.
\item{GAT~\cite{velivckovic2017graph}:} A wildly used attention-based GCN model in the homogeneous graph that dynamically assigns weights for neighbors.
\item{HAN~\cite{wang2019heterogeneous}:} A state-of-the-art GCN-based network embedding model for HINs. We adapt the model for recommendation task and keep the original meta-structure design where the meta-structure links two homogeneous types of nodes.
\item{FMG \cite{zhao2017meta}:} The state-of-the-art model leveraging meta-structures in HIN recommendation. FMG combines matrix factorization with factorization machine to relieve data sparsity.
\item{GEMS-fix:} Instead of searching all possible meta-structures, we preset some hand-crafted meta-structures according to previous papers to exam the performance of our model. The adopted meta-structures are shown in Table \ref{tab:metapath}.
\end{itemize}

\begin{table}[t]
\small
\begin{center}
\caption{Hand-crafted meta-struct. designs from previous papers.}
\begin{tabular}{c|c}
\Xhline{1pt}
  Dataset&  Meta-Structures\\
\Xhline{0.8pt}
Yelp & U-B, U-U-B, U-B-U-B, U-B-A-B, U-B-I-B \\
Douban Movie  & U-M, U-M-U-M, U-G-U-M, U-M-A-M, U-M-T-M \\
Amazon      & U-I, U-I-U-I, U-I-V-I, U-I-B-I, U-I-C-I\\
\Xhline{1pt}
\end{tabular}
\label{tab:metapath}
\end{center}
\vspace{-0.1in}
\end{table}

\subsection{Experimental Setup and Reproducibility}

For evaluation, we divide each dataset into three parts for training, validating and testing with 8:1:1 ratio. Since it is inefficient to rank the test items with the entire item set, for each train/test record, we randomly sample 4/100 negative items based on the popularity to evaluate the models. Then we adopt three commonly used performance metrics as \emph{Hit Ratio at Rank K} (HR@K), \emph{Mean Reciprocal Rank at Rank K} (MRR@K) and \emph{Normalized Discounted Cumulative Gain at Rank K} (NDCG@K). HR@K accounts for whether test items are present in the top-k list, and MRR@K and NDCG@K measure the ranking positions of test positive items.

For the baseline models, we reference the implementations released by the authors and change the loss function into a margin-based ranking loss for recommendation purpose. For the Yelp and Amazon dataset, all the GCN based models like PinSAGE, GAT, HAN and GEMS adopt MF pre-trained embeddings as feature inputs. Considering the number of interactions is huge in Douban dataset, we let these models learn their own feature inputs to avoid limiting their representational ability. 
We fix the dimensions of embeddings for all evaluated models at 64, and tune the learning rate and regularization parameters to optimal for each model by grid search. Specifically, we adopt the ADAM optimizer with learning rate decay for fine-grained results. For our GEMS, we set the population size as 20, which means 20 individuals in each generation. Each individual contains 5 genes, that are 5 corresponding meta-structures. The mutation probability is set to 0.6 in the first few generations, then decreases to 0.3. For each mutation, we set the equal probability for being complex or simple to give more freedom to the searching process in a random walk manner. 

The implementation code of our model is available at \url{https://github.com/0oshowero0/GEMS}.

\subsection{Overall Performance Analysis}

\begin{table*}[t]
    \small
	\centering
	\caption{Performance comparison with baseline models, where ($\ast $) indicates p<0.01 significance over best baseline.}
	\begin{tabular}{m{1.5cm}<{\centering} |m{0.6cm}<{\centering} m{0.7cm}<{\centering} m{0.8cm}<{\centering} m{0.8cm}<{\centering} m{0.8cm}<{\centering} |m{0.7cm}<{\centering} m{0.8cm}<{\centering} m{0.8cm}<{\centering} m{0.8cm}<{\centering} m{0.8cm}<{\centering} |m{0.7cm}<{\centering} m{0.7cm}<{\centering} m{0.8cm}<{\centering} m{0.8cm}<{\centering} m{0.8cm}<{\centering}}
	    \hline
		\multirow{2}{*}{Method}&\multicolumn{5}{c|}{Yelp}&\multicolumn{5}{c|}{Douban Movie}&\multicolumn{5}{c}{Amazon}\cr\cline{2-16} 
		&\footnotesize HR3  & \footnotesize MRR10  & \footnotesize NDCG10 & \footnotesize MRR50 & \footnotesize NDCG50& \footnotesize HR3  & \footnotesize MRR10  & \footnotesize NDCG10 & \footnotesize MRR50 & \footnotesize NDCG50& \footnotesize HR3  & \footnotesize MRR10  & \footnotesize NDCG10 & \footnotesize MRR50 & \footnotesize NDCG50\\ \hline
		NMF&0.1321&0.1172&0.1692&0.1431&0.2718&0.1359&0.1254&0.1745&0.1430&0.2763&0.1249&0.1197&0.1640&0.1408&0.2739\\
        BMF&0.1299&0.1148&0.1671&0.1366&0.2696&0.1358&0.1211&0.1726&0.1447&0.2844&0.1380&0.1223&0.1739&0.1448&0.2810\\
        Metapath MF&0.1255&0.1124&0.1625&0.1343&0.2670&0.1274&0.1133&0.1634&0.1370&0.2753&0.1301&0.1144&0.1639&0.1361&0.2674\\
        SVD++&0.1463&0.1316&0.1733&0.1521&0.2811&0.1394&0.1241&0.1767&0.1482&0.2845&0.1505&0.1279&0.1773&0.1494&0.2876\\
        \hline
        PinSAGE &0.1694&0.1456&0.2033&0.1662&0.3005&0.1463&0.1287&0.1808&0.1517&0.2902&0.1469&0.1285&0.1801&0.1503&0.2835\\
        GAT &0.1706&0.1459&0.2020&0.1665&0.2997&0.1345&0.1189&0.1683&0.1418&0.2786&0.1510&0.1349&0.1883&0.1566&0.2908\\
        \hline
        HAN &0.1674&0.1459&0.2047&0.1675&0.3056&0.1496&0.1322&0.1854&0.1552&0.2945&0.1525&0.1348&0.1900&0.1572&0.2953\\
        FMG&0.1765&0.1486&0.2062&0.1675&0.2957&0.1510&{0.1336}&0.1800&0.1572&0.2901&0.1532&0.1313&0.1877&0.1555&0.2910\\
        \hline
        GEMS-fix&0.1611&0.1405&0.1967&0.1621&0.2981&0.1437&0.1265&0.1779&0.1495&0.2874&0.1505&0.1322&0.1861&0.1545&0.2912\\
        GEMS&{\bf0.1776}&{\bf0.1529$ \ast$}&{\bf0.2106$ \ast$}&{\bf0.1737$\ast $}&{\bf0.3092}&
        {\bf0.1542$\ast $}&{\bf0.1355$\ast $}&{\bf0.1888$\ast$}&{\bf{0.1584}}&{\bf0.2976$\ast $}&{\bf0.1770$\ast $}&{\bf0.1540$\ast $}&{\bf0.2111$\ast $}&{\bf0.1759$\ast $}&{\bf0.3133$\ast $}\\
        \hline
	\end{tabular}
	\label{tab:result}
\end{table*}

The main experiment results of comparing with the baselines across the three datasets are reported in Table \ref{tab:result}. From the results, we have the following observations and conclusions.
\begin{itemize}[leftmargin=10px]
\item  In all the three datasets, our proposed GEMS model constantly outperforms the baselines on all evaluation metrics. These results demonstrate the GEMS model can find a more useful meta-structure structure to leverage heterogeneous information better than baseline models. 

\item GEMS with the hand-crafted meta-structures does not catch up with the state-of-the-art model like HAN and FMG. However, with meta-structure search, the GEMS model greatly outperforms the best baseline method. This phenomenon confirms the importance of meta-structure design in HIN recommendation.

\item Compare GEMS with GEMS-fix, we can observe that GEMS outperforms GEMS-fix by 11.7\%, 8.8\% on HR@3 and NDCG@10 on average. GEMS-fix adapts commonly used hand-crafted meta-structures for recommendation, while it greatly falls behind the searched meta-structures of GEMS. This phenomenon demonstrates that hand-crafted meta-structures are not optimal for recommendation.


\item For HINs with different relationships and densities, GEMS surpasses all state-of-the-art-models. In Douban Movie dataset, the link density of target nodes is over 6 times bigger than Yelp dataset, which prefers dedicate designed models like HAN or FMG. However, the GEMS model still outperforms all baselines, demonstrating its adaptability over different datasets.

\end{itemize}

Above analysis demonstrate the effectiveness of our proposed GEMS model, which constantly achieves better performance on different recommendation scenarios.

\subsection{Training Efficiency Analysis}

\begin{figure*}[t]
 \centering
    \subfigure[Yelp dataset]{
        \label{fig:perf_yelp}
        \includegraphics[width=0.3\textwidth]{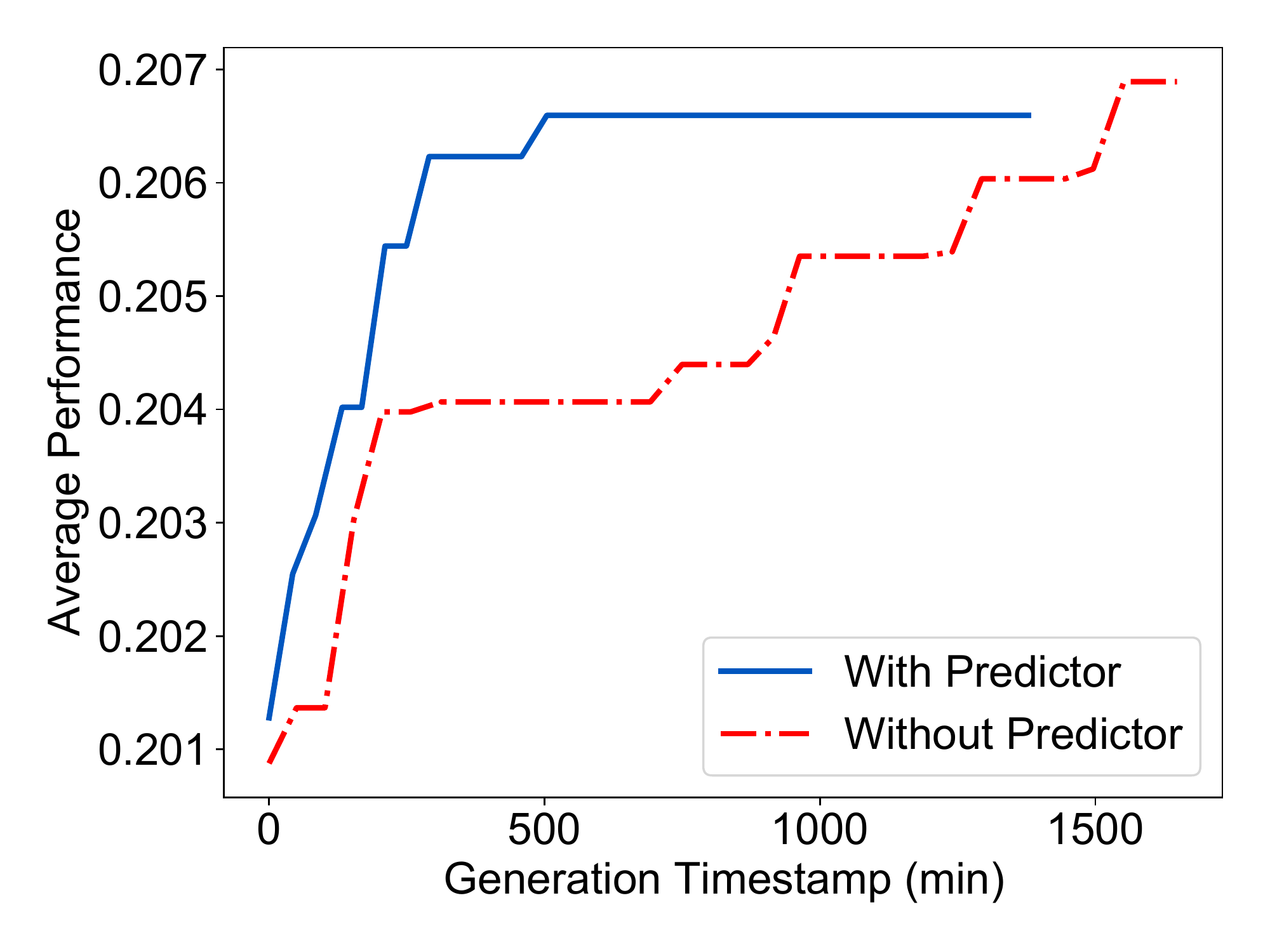}}
    \subfigure[Douban Movie dataset]{
        \label{fig:perf_douban}
        \includegraphics[width=0.3\textwidth]{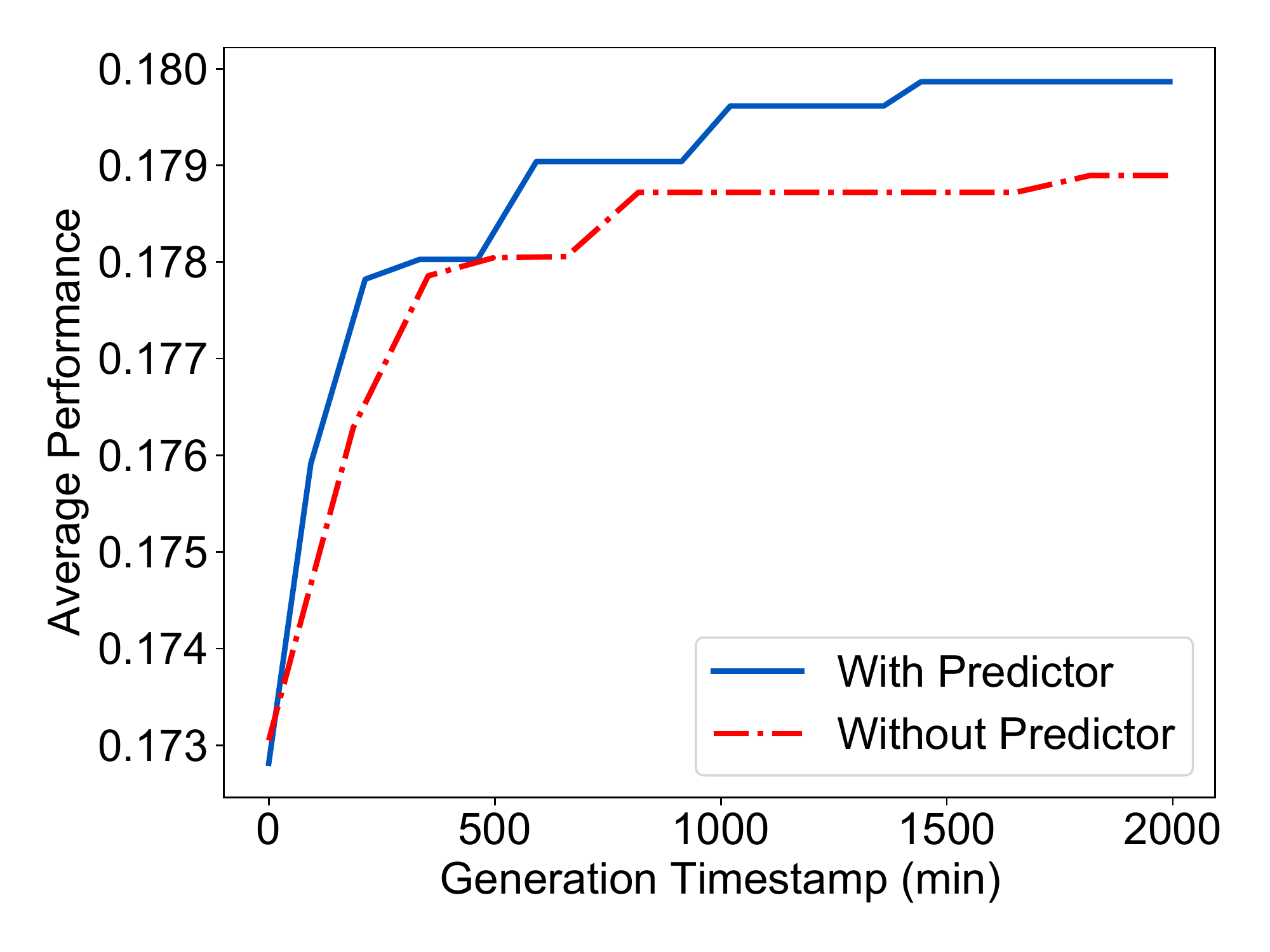}}
    \subfigure[Amazon dataset]{
        \label{fig:perf_amazon}
        \includegraphics[width=0.3\textwidth]{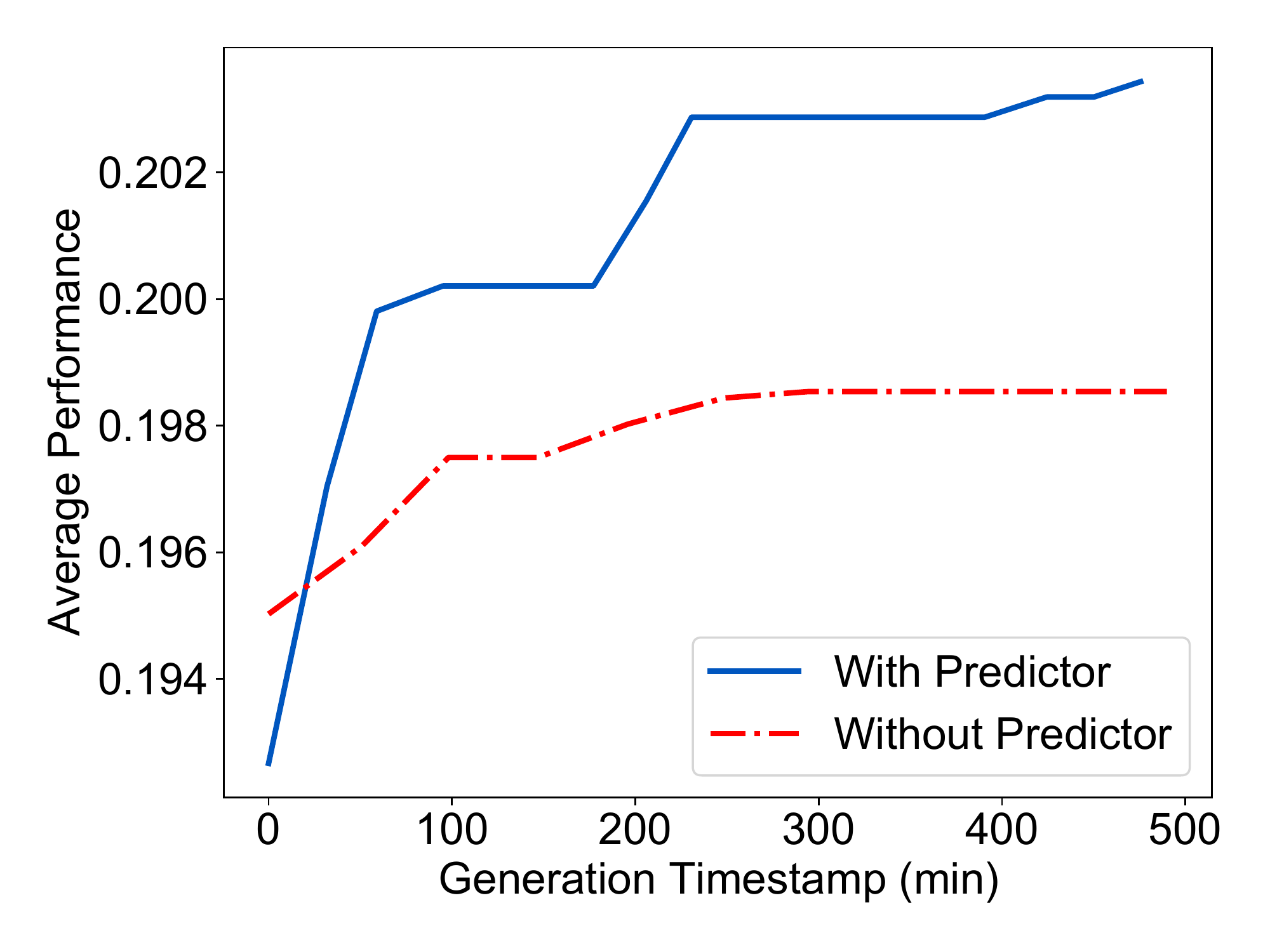}}
    \vspace{-0.1in}
 \caption{The time efficiency analysis of GEMS.}
 \vspace{-0.1in}
 \label{fig:perf-time}
\end{figure*}

To further understand the efficiency of GEMS, we move forward to analyze the performance-time curve on recommendation tasks. An important note is that the search process only exists in the first run. After we found the effective meta-structures, we can omit the search process by directly set the optimized meta-structures. Since GEMS only use one-layer GCN, the inference time should be faster than traditional multiple-layer GCN baselines.

Figure \ref{fig:perf-time} shows the performance evolution of model performance with time on three different datasets. We record the average performance (NDCG@10) of each generation, and choose the maximum performance before a given timestamp. In that case, we can clearly capture the performance evolution of the whole population rather than some outstanding individuals. The initial meta-structure is the direct connection of target interactions for Yelp and Douban Movie dataset. However, for Amazon dataset, initializing by direct interaction will greatly limit the searching process due to the limited relations. To break the symmetry we randomly initialize the meta-structures, which results in the difference at the beginning. 

For Yelp dataset, we can observe a clear convergence process for both situations. The performance gain on Yelp dataset is over $2.8\%$. For Douban and Amazon dataset, the GEMS model achieves $3.9\%$ and $5.6\%$ performance gain accordingly. Considering each recorded performance is averaged for the whole generation, we can draw the conclusion that the search process of GEMS is effective. 

We also compare the performance of GEMS in both with predictor and without predictor to exam whether the predictor can help us to explore more meaningful meta-structures in the same period of time. For Yelp dataset, GEMS with predictor achieves a relatively higher performance within 500 minutes, while the model without predictor takes nearly 1500 minutes to achieve the same level. The meta-structure predictor has great potential to boost the convergence of GEMS. For Douban Movie dataset, GEMS with predictor achieves $3.9\%$ performance gain compared with the initial state. Douban Movie shares the most complex relations, which require the model to explore more meta-structures. GEMS without predictor does not achieve better performance than the model with predictor one during the whole time window, which indicates the possible performance gain due to the meta-structure predictor. Similar behavior exists in Amazon dataset. In conclusion, the meta-structure predictor is helpful in reducing the required training time as well as improving performance.

\subsection{In-depth Meta-Structure Evaluation}

\begin{figure}[t]
 \centering
    \subfigure[Yelp dataset]{
        \label{fig:gene_yelp}
        \includegraphics[width=0.44\textwidth]{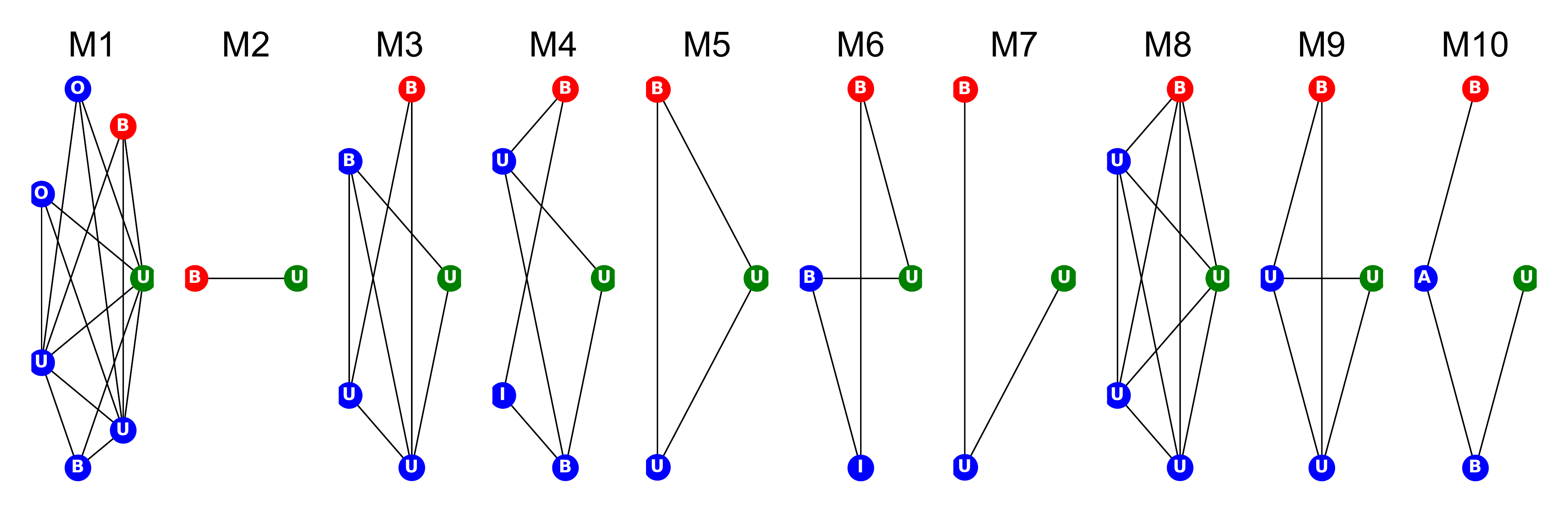}}
    \vspace{-0.1in}
    \subfigure[Douban Movie dataset]{
        \label{fig:gene_douban}
        \includegraphics[width=0.44\textwidth]{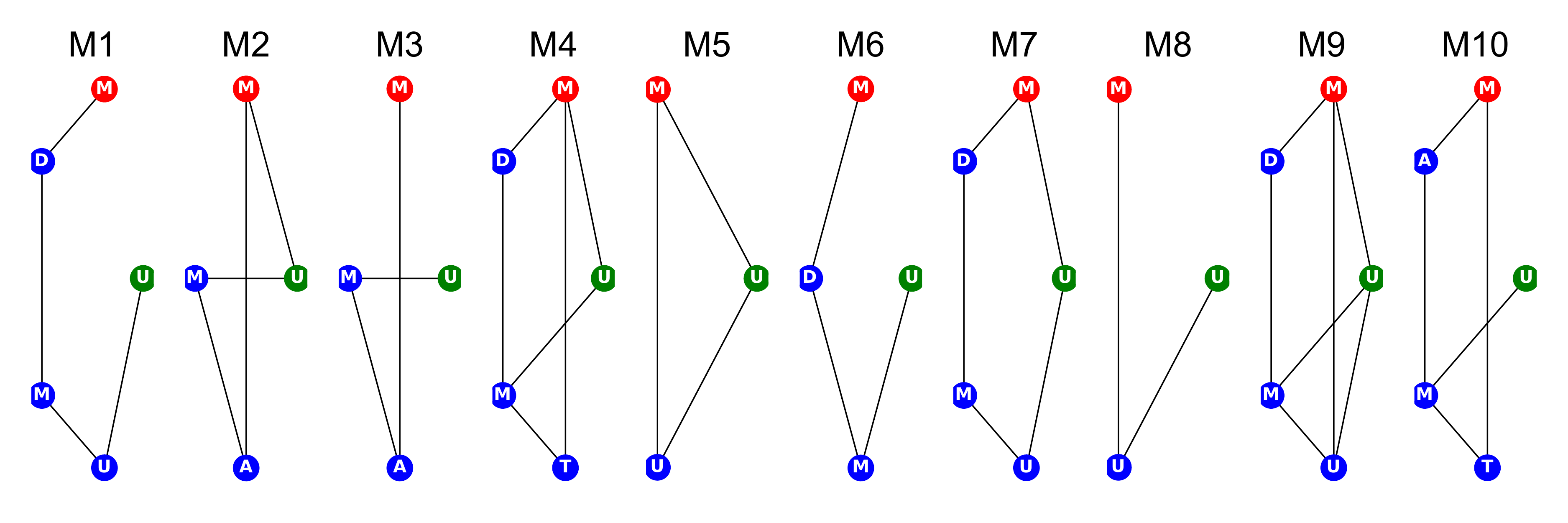}}
    \vspace{-0.1in}
    \subfigure[Amazon dataset]{
        \label{fig:gene_amazon}
        \includegraphics[width=0.44\textwidth]{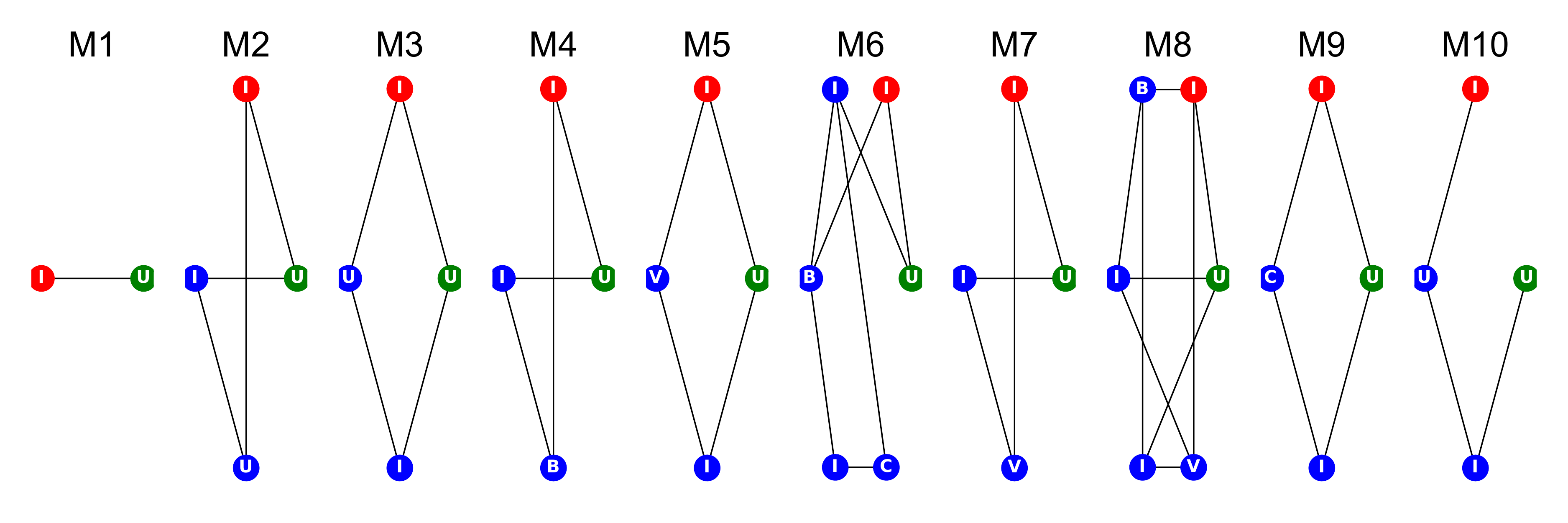}}
 \caption{Most important meta-structures identified by GEMS.}
 \label{fig:genes}
   \vspace{-0.2in}
\end{figure}

In order to investigate whether the proposed GEMS model can find new meta-structures that are meaningful for recommendation tasks, we exam the most frequent meta-structures in the last 5 generations of GEMS to analyze the physical meaning.

As shown in Figure \ref{fig:genes}, we demonstrate the top 10 meta-structures each dataset. The green node and red node are target nodes of the meta-structure. These genes accounts for $39.2\%,25.6\%,44.5\%$ of all 500 meta-structures during the last 5 generations. We labeled each meta-structures from M1 to M10 for short. Some of the searched meta-structures are already known by us, while others are never shown in recommender system designs. Thus, we divided the meta-structures into two groups in Table \ref{tab:metagraphs}.

In Yelp dataset, we already know the direct interactions and social-related meta-structures are beneficial for recommendation tasks. For example, M2 depicts the direct relationship between users and businesses, M5 and M7 depict the behavior of common interest between friends. Besides, the collaborative filtering meta-structure M10 often appears in meta-structure based recommendation models. At the same time, new meta-structures can also facilitate recommendations based on their semantics. M6 emphasizes the importance of locality, where users tend to visit local businesses in the same city. M3 and M9 depict enhanced social groups that the same business visited by more than one friend is also attractive for the target customer. What's more, M4 delivers a more complex relationship that a user tends to patronize the local business that her friend has been to. These searched meta-structures show that social relations and locality play an important role in Yelp dataset. 

For movie recommendation in Douban Movie, we can find many collaborative filtering meta-structures like M2, M3, and M6. According to the traditional understanding, users tend to appreciate new movies of the same actor or the same director. We can also find similar social relations as Yelp in M5, M8, where friends tend to share the same interest in movies. New meta-structures like M4 and M10 are two enhanced versions of collaborative filtering. For example, users prefer new movies with the same director and the same type with old ones according to M4. Besides, M1 and M7 combine social relation with collaborative filtering, where the collaborative filtering path links the friend of the target user.

In Amazon dataset, user-user relationship does not exist. Users can only connect to item nodes, generating a traditional recommendation scenario. The resulting meta-structures show that most of the meta-structures can be regarded as a simple modification on collaborative filtering meta-structures. It shows that the GEMS can also work well in traditional recommendation scenarios. 

From these results and analysis, we can confirm the effectiveness of the searched meta-structures by GEMS. Besides, these meta-structures can also provide new inspirations on recommender system design. First, there are only 7 meta-structures among these 30 meta-structures, which confirms the effectiveness of meta-structures for HIN recommendation. Newly designed recommendation models should adopt meta-structures to better leverage the complex relations of HINs. Another important insight is that there is no perfect design paradigm that suits all recommendation scenarios. For Yelp users, social factors are very important. While for the Douban Movie or Amazon users, they often pay more attention to the item itself. During the design of meta-structures, we need to consider the core feature of scenarios for a better recommendation.


\begin{table}[t]
\small
\begin{center}
\caption{Hand-crafted and newly searched meta-structures.}
\begin{tabular}{c|c|c}
\Xhline{1pt}
  Dataset&  Hand-crafted Meta-Struct. &Newly Searched Meta-Struct. \\
\Xhline{0.8pt}
Yelp &M2, M5, M7, M10 &M1, M3, M4 ,M6, M8, M9 \\
Douban  &M2, M3, M5, M6, M8 &M1, M4, M7, M9, M10\\
Amazon   &M1-M5, M7, M9, M10 &M6, M8\\
\Xhline{1pt}
\end{tabular}
\label{tab:metagraphs}
\vspace{-0.2in}
\end{center}
\end{table}

\subsection{Effectiveness of Meta-Structure Predictor}

\begin{figure}[t]
 \centering
    \vspace{-0.1in}
    \subfigure[Training result on Amazon dataset]{
        \label{fig:pred_tr_amazon}
        \includegraphics[width=0.245\textwidth]{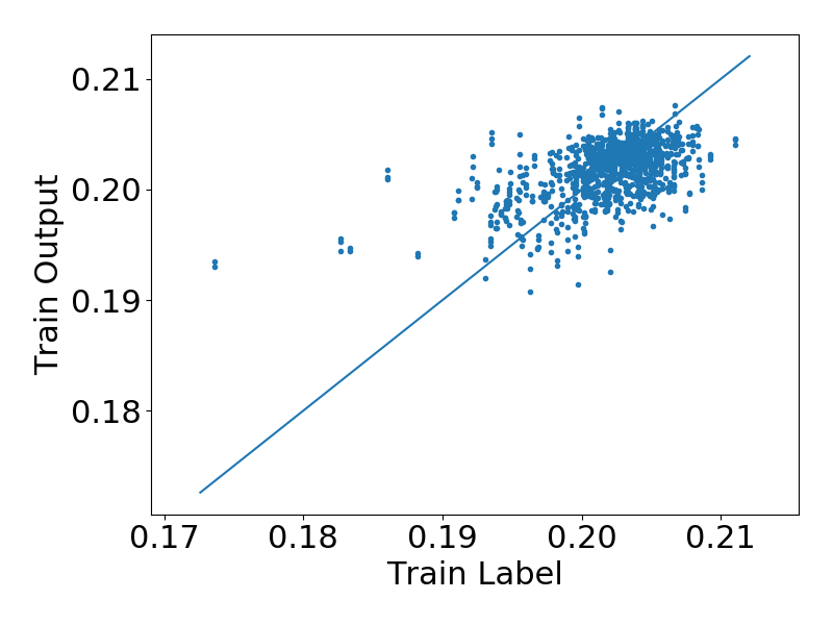}}%
    \subfigure[Testing result on Amazon dataset]{
        \label{fig:pred_test_amazon}
        \includegraphics[width=0.245\textwidth]{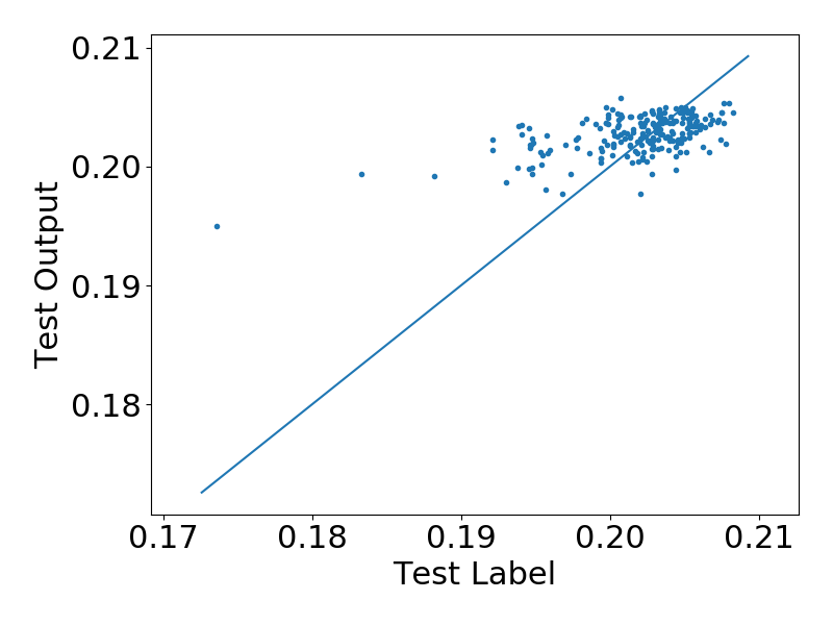}}
 \caption{Effectiveness of meta-structure predictor.}
 \vspace{-0.1in}
 \label{fig:predictor}
\end{figure}

To further evaluate the effectiveness of the proposed meta-structure predictor, we train meta-structure predictors on the historical performance of each individual, then compare their performance on both training set and test set. Figure \ref{fig:predictor} shows the relation between the true label and predictor output on Amazon dataset. Ideally, the predicted results should be exactly the same as the true label. However, since the predictor evaluates the performance at individual level, it is not able to hard remember the performance of meta-structures, which brings more difficulties during inference. We need to note that the absolute output is not a key factor to evaluate the predictor. Since the meta-structure predictor is designed to select promising individuals, we only expect the relative performance is correct. In that case, the promising individuals will still surpass worse ones and sent to be evaluated on real recommendation tasks. To evaluate the ranking performance, we further exam the Spearman ranking correlations between the predictor output and true labels as shown in Table \ref{tab:spearman}. We can find that the ranking performance of meta-structure predictor is promising for unseen data on the test set. Compared with other automated predictors like \cite{liu2018progressive}, our meta-structure predictor achieves reasonable performance on model selection.

\begin{table}[t]
\small
\begin{center}
\caption{Spearman correlations of meta-structure predictors.}
\begin{tabular}{c|c|c}
\Xhline{1pt}
  Dataset&  Training set & Test set\\
\Xhline{0.8pt}
Yelp & 0.5499&0.3426 \\
Douban Movie  & 0.4517&0.4946 \\
Amazon    &0.5313  &0.4160\\
\Xhline{1pt}
\end{tabular}
\label{tab:spearman}
\vspace{-0.1in}
\end{center}
\end{table}

\subsection{Parameter Sensitivity}
We investigate the sensitivity of parameters for GEMS model on regularization parameters and embedding dimension. Experiments are implemented on Yelp dataset.

\begin{figure}[t]
 \centering
    \subfigure[Regularization]{
         \label{fig:lamb}
         \includegraphics[width=.23\textwidth]{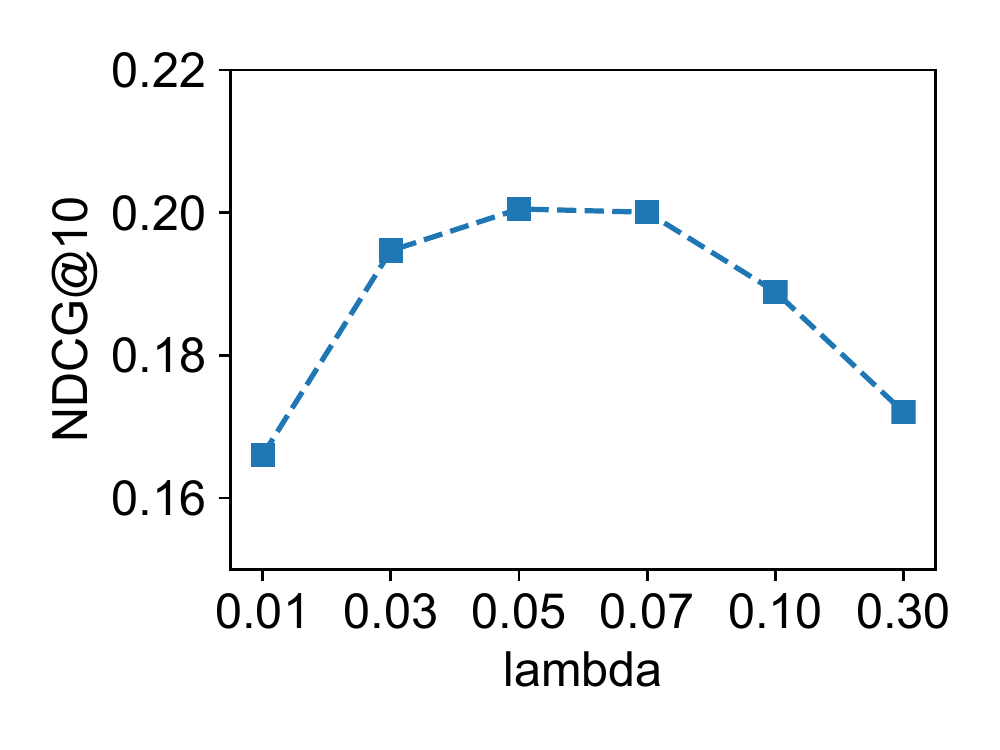}}
    \subfigure[Embedding size]{
        \label{fig:emb}
        \includegraphics[width=.23\textwidth]{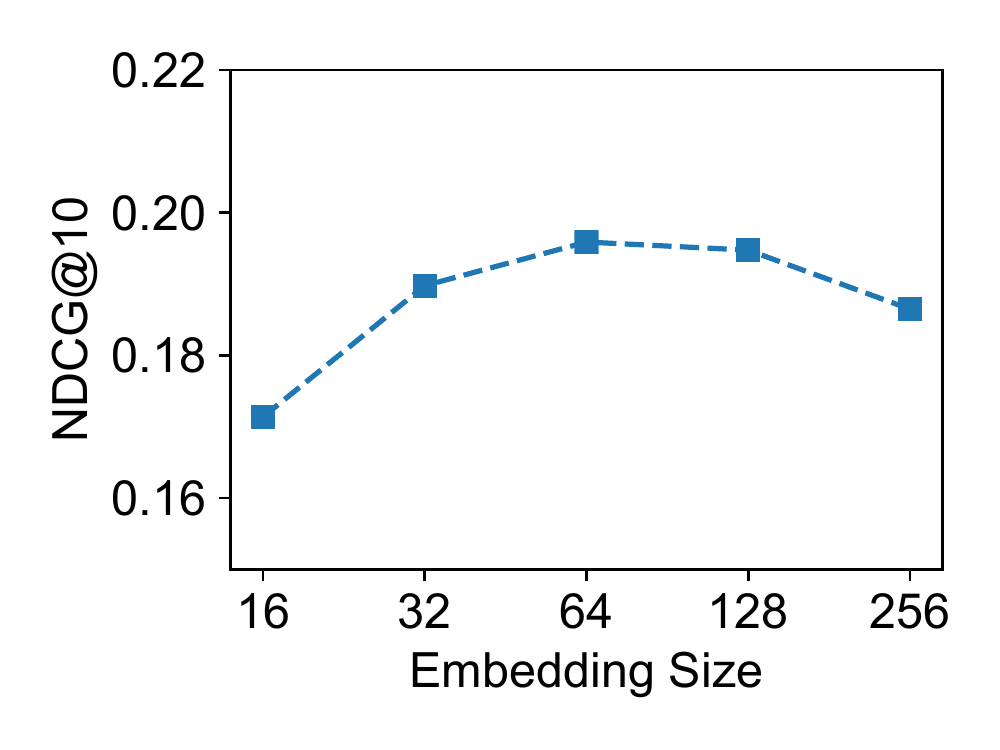}}
 \caption{Parameter sensitivity on Yelp dataset.}
 \label{fig:param}
  \vspace{-0.2in}
\end{figure}

From Figure \ref{fig:param}, we can observe that the performance varies with regularization parameters (lambda) and the embedding dimensions, but the NDCG@10 consistently remains above 0.16. Specifically, the optimal regularization parameter for GEMS is 0.05. In addition, the NDCG@10 consistently increases from $0.1714$ to $0.1959$ when the embedding dimensions increase from $16$ to $64$, and it starts to go down when the embedding dimensions are 128. This phenomenon validates the hyper-parameter choices during experiment section. 
 

\section{Related Work}\label{sec:Related}

\textbf{Graph convolutional networks:} 
The idea of GCNs is to design a spatial in-variant aggregate function to generate node embeddings by aggregating features from their local neighbourhood~\cite{kipf2016semi}. In recommendation systems, user-item connections naturally form a network structure, which suits the GCN based models. PinSage \cite{ying2018graph} constructs GCN on item graph to learns item-to-item similarity, and DiffNet \cite{wu2019neural} performs social diffusion process on user-user graph by GCN. However, using a homogeneous graph greatly limits the expressiveness of GCN model. In that case, heterogeneous information networks are becoming the mainstream setup of GCN.

\noindent\textbf{Heterogeneous information networks:} Heterogeneous information network \cite{sun2011pathsim} is a well-established framework that contains different types of nodes and relations, which processes much more semantic information than the traditional homogeneous network. To effectively guide information propagation in HINs, meta-path based models like PathSim \cite{sun2011pathsim} have been proposed. Meta-path based models demonstrate promising results on HIN setting, which stimulates many data mining tasks like recommendation \cite{shi2015semantic}, similarity search and classification \cite{bangcharoensap2016transductive} to leveraging the power of HINs. Except for meta-path based model, the meta-structure has also become a new topic in HINs. In recommendation area, social factors are important for a success recommendation \cite{xu2019think}, which requires HINs to improve the recommendation quality. RecoGCN \cite{xu2019relation} proposes a meta-path based attention GCN in agent-initiated social e-commerce scenario. FMG \cite{zhao2017meta} adopts a MF+FM schema to leverage meta-structures for recommendation. As a more general term, we use meta-structure to represent both the two concept. 

\noindent\textbf{Automated machine learning:} 
AutoML is a broad concept that applies to all automated models that try to take the place of humans on identifying proper configurations in machine learning concept. Automated model selection \cite{xie2017genetic} is a common examples of AutoML. In this paper, the proposed GEMS leverages genetic algorithm to find meaningful meta-structures, which is the first try on automated meta-structure searching for HIN recommendation.

\section{Conclusion}\label{sec:Conclusion}

In this paper, we proposed a novel model leveraging automated machine learning paradigm to search promising meta-structures for HIN recommendation. To effectively explore possible meta-structures, we carefully designed the search space of the problem which boosts the searching efficiency. During recommendation, we proposed a multi-view GCN armed by attention mechanism to fuse different semantic information guided by meta-structures. Extensive experiments demonstrate the performance gain in GEMS, where the optimized meta-structures also shed light on sophisticated recommendation model design in HIN scenario. Important future works will be conducted on a more dedicated downstream scorer function to generate better recommendations.

 
\begin{acks}
This work was supported in part by The National Key Research and Development Program of China under grant 2018YFB1800804, the National Nature Science Foundation of China under U1936217,  61971267, 61972223, 61941117, 61861136003, Beijing Natural Science Foundation under L182038, Beijing National Research Center for Information Science and Technology under 20031887521, and research fund of Tsinghua University - Tencent Joint Laboratory for Internet Innovation Technology.
\end{acks}

\bibliographystyle{ACM-Reference-Format}
\bibliography{references}
\clearpage
\end{document}